\documentclass[aps, prl, twocolumn, superscriptaddress, tightenlines, longbibliography]{revtex4-2}

\usepackage{graphicx}
\usepackage{amsmath}
\usepackage{booktabs}
\usepackage{subfigure}
\usepackage{epstopdf}
\usepackage{bm}
\usepackage{natbib}
\usepackage[colorlinks=true,linkcolor=blue,citecolor=blue,urlcolor=blue]{hyperref}
\usepackage{pdfpages}
\makeatletter
\AtBeginDocument{\let\LS@rot\@undefined}
\makeatother

\begin{document}

	\title{Electromagnetically-Induced Transparency Bridges \\ Disconnected Light-Harvesting Networks}
	
	\author{Jun Wang}
\affiliation{School of Physics and Astronomy, Applied Optics Beijing Area Major Laboratory, Beijing Normal University, Beijing 100875, China}
\affiliation{Key Laboratory of Multiscale Spin Physics, Ministry of Education, Beijing Normal University, Beijing 100875, China}

	\author{Rui Li}
\affiliation{School of Physics and Astronomy, Applied Optics Beijing Area Major Laboratory, Beijing Normal University, Beijing 100875, China}
\affiliation{Key Laboratory of Multiscale Spin Physics, Ministry of Education, Beijing Normal University, Beijing 100875, China}
	
	\author{Yi Li}
\affiliation{School of Physics and Astronomy, Applied Optics Beijing Area Major Laboratory, Beijing Normal University, Beijing 100875, China}
\affiliation{Key Laboratory of Multiscale Spin Physics, Ministry of Education, Beijing Normal University, Beijing 100875, China}

	\author{Kai-Ya Zhang}
\affiliation{School of Physics and Astronomy, Applied Optics Beijing Area Major Laboratory, Beijing Normal University, Beijing 100875, China}
\affiliation{Key Laboratory of Multiscale Spin Physics, Ministry of Education, Beijing Normal University, Beijing 100875, China}

	\author{Qing Ai}
	\email{aiqing@bnu.edu.cn}
\affiliation{School of Physics and Astronomy, Applied Optics Beijing Area Major Laboratory, Beijing Normal University, Beijing 100875, China}
\affiliation{Key Laboratory of Multiscale Spin Physics, Ministry of Education, Beijing Normal University, Beijing 100875, China}
	
	
	\date{\today}

	\begin{abstract}
		The energy-transfer efficiency of the natural photosynthesis system seems to be perfectly optimized during the evolution for millions of years. However, how to enhance the efficiency in the artificial light-harvesting systems is still unclear. In this paper, we investigate the energy-transfer process in the photosystem I (PSI). When there is no effective coupling between the outer antenna (OA) and the reaction center (RC), the two light-harvesting networks are disconnected and thus the energy transfer is inefficient. In order to repair these disconnected networks, we introduce a bridge with three sites between them. We find that by modulating the level structure of the 3-site bridge to be resonant, the energy transfer via the dark state will be enhanced and even outperform the original PSI. Our discoveries may shed light on the designing mechanism of artificial light-harvesting systems.
	\end{abstract}
	
	\maketitle
	
\section{Introduction}
	

The demand for sustainable energy solutions has increasingly turned to the sun, an inexhaustible source of power that holds the promise of resolving our ongoing energy crises. In natural photosynthesis, green plants and other photosynthetic organisms perform excellent efficiency of converting and storing solar energy. This biological process occurs within the chloroplasts' thylakoid membranes, whose structure has been optimized during the evolution for million of years. Inspired by the natural photosynthesis, artificial photosynthetic systems have been fervently developed to mimic these biological processes to harness solar energy effectively \cite{Rybtchinski2004JACS, Fukuzumi2008PCCP, Reece2011Science, Alibabaei2015PNAS, Turan2016NC, Chowdhury2018NC, Sokol2018NE, Zhou2019NC, Dogutan2019ACR, Wang2022CSR, Wang2024NC, Mori2025NC}.

At the heart of natural photosynthesis are photosystem I (PSI) \cite{Mazor2017natpl} and photosystem II \cite{Wei2016nature} (PSII). These multi-subunit complexes are embedded within the thylakoid membranes and are responsible for light harvesting \cite{Croce2020science}. The natural efficiency of these systems is largely due to their light-harvesting antennae, which capture a broad spectrum of wavelengths using multiple chlorophyll molecules. This large cross-sectional area of chlorophyll does not only capture sunlight efficiently but also facilitates rapid energy transfer to the reaction center (RC) \cite{Nelson2015arb, Qin2015science, Pan2018science, Iwai2018natpl, Qin2019natpl, Stirbet2020aob, Caspy2020natpl, Nagao2020nc, Harris2023nc}. To enhance the efficiency of artificial systems, much effort has been paid to the study of the mechanisms of PSI and PSII, aiming to replicate and optimize these processes in synthetic setups \cite{wilde1995pc, stockel2006pp, albus2010pc, liu2012pc, shen2017pp, nellaepalli2018nc, su2019natpl, Suga2019natpl, garab2023joeb, you2023nature, feng2023sciadv, zhang2024natpl}.

However, in the artificial photosynthetic systems, the outer antenna (OA) might be far away from the RC, leading to weak coupling between them. A direct solution is to insert a bridge to connect the OA and the RC. The bridge should consist of multiple sites in order to cross over a sufficient long distance. Thus, it is crucial to optimize the bridge to reduce the dissipation of the intermediate site in the bridge. The electromagnetically-induced transparency (EIT) \cite{fleischhauer2005rmp, liu2016pra, gu2016pra, wang2018pra, Wang2018dark, Wang2020NP, Andersson2020PRL, McDonnell2022PRL, Kim2023PRL} offers an intriguing possibility for this problem. The EIT is a phenomenon that the population on the lossy intermediate level of a three-level system is sufficiently suppressed when the energy levels satisfy the two-photon resonance. The state without population on the intermediate level is named as the ``dark state", which can be used to improve the energy transfer efficiency \cite{dong2012light}. By applying the EIT to the energy-transfer processes in the artificial photosynthetic system, it is possible to bridge the OA and the RC in such way that energy loss through spontaneous emission at intermediate sites is minimized, and the overall efficiency of the artificial light-harvesting system is increased.

In this paper, we  investigate the energy-transfer process in an inefficient light-harvesting system including two disconnected networks. Because the OA is far away from the RC, their interaction is weak and the energy can not be effectively transferred from the OA to the RC. In order to repair the network, we introduce a bridge with three sites between the OA and the RC. When both the single-photon and two-photon resonance are satisfied, the decay via the intermediate site is suppressed during the energy-transfer process through the dark state. Furthermore, when the sites energies of the the 3-site bridge are the same as in the natural PSI, the energy-transfer efficiency can be further optimized. However, when there is two-photon resonance but large single-photon detuning, the efficiency is slightly decreased compared with the natural PSI case, which implies that the natural PSI has been sufficiently optimized during the evolution in the past millions of years. Interestingly, the energy-transfer efficiency can be further optimized as compared to the natural PSI when the single-photon resonance is achieved. This bridge structure also works well in an artificial light-harvesting system. In this regard, we can repair two disconnected light-harvesting networks by introducing a bridge assisted by the EIT.

\begin{figure}[!ht]
	\centering
	\includegraphics[width=8.5cm]{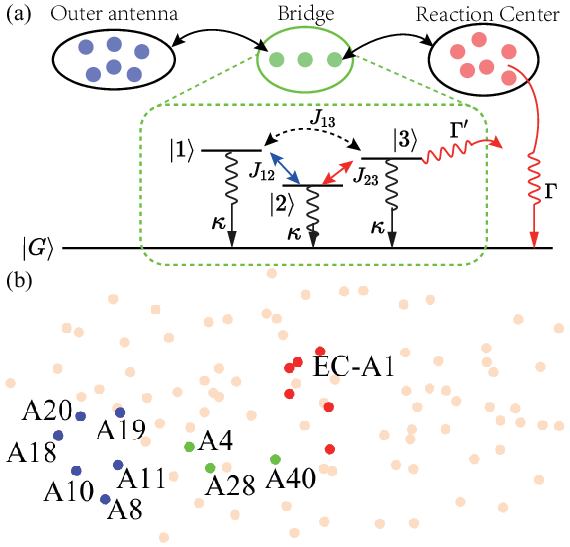}
	\caption{(a) Schematic of disconnected light-harvesting networks repaired by an EIT bridge. Each dot represents a chlorophyll with a ground state and an excited state. A bridge with three sites is inserted between the OA and the RC to connect them. Under the single-excitation condition, $|m\rangle$ represents the excitation on site $m$, while the other sites are in the ground state. $\kappa$ is the spontaneous-emission rate, and $\Gamma$ represents the charge-separation rate in the RC. For the simplified 3-site bridge model, $\Gamma'$ is the effective rate of site 3 in the bridge to the RC. (b) The position of the Mg atoms of the chlorophyll molecules in PSI. Blue, green and red dots respectively represent the OA, the bridge and the RC sites. The gray dots are the other sites in the PSI.}\label{fig:scheme}
\end{figure}

\section{Model}


We consider a light-harvesting system described by the density matrix $\rho$ under the single-excitation condition. The system contains two networks, i.e., the OA and the RC, and a bridge between them, as shown in Fig.~\ref{fig:scheme}(a). Each site in Fig.~\ref{fig:scheme}(a) represents a chlorophyll modeled as a two-level system. The coherent part of the dynamics of the system can be modeled by the $N$-qubit Hamiltonian describing the coherent exchange of excitations between different sites, i.e.,
\begin{align}
	H=\sum_{n=1}^{N} E_n|n\rangle\langle n|+\sum_{m\neq n}J_{mn}|m\rangle\langle n|,
\end{align}
where $\hbar=1$, $|n\rangle$ denotes a single excitation in site $n$, $E_n$ is the local site energy, $J_{mn}$ is the hopping rate of the excitation between sites $m$ and $n$ induced by the dipole-dipole interaction. Please refer to Ref.~\cite{Damjanovic2002jpcb} for more details. The Hamiltonian $H$ can be diagonalized as $H=\sum_{k}\varepsilon_{k}|\varepsilon_{k}\rangle\langle \varepsilon_{k}|$ with the eigen energy $\varepsilon_{k}$ and the eigen state $|\varepsilon_{k}\rangle$.


We represent the biochemical reaction in the RC by the decay of the excitation on the final site labeled by $|N\rangle$ to the ground state $|G\rangle$, which is modeled by the Lindblad term $\mathcal{L}_{\textrm{rc}}(\rho)=\Gamma\mathcal{D}(|G\rangle\langle N|)\rho$, with $\Gamma$ being the biochemical-reaction rate and $\mathcal{D}(A)\rho=A\rho A^{\dag}-(\rho A^{\dag}A+A^{\dag}A\rho)/2$. In addition, the spontaneous emission for the excitation on each site to the ground state $|G\rangle$ can be modeled by the local Lindblad term $\mathcal{L}_{\textrm{em}}(\rho)=\sum_{n}\kappa\mathcal{D}(|G\rangle\langle n|)\rho$, with $\kappa$ being the spontaneous-emission rate.

The coherent modified Redfield theory (CMRT) and its generalization \cite{Yang2002CP, Hwangfu2015CP, chang2015jcp, de2017rmp, jang2018rmp, tao2020sb} have been widely used to describe the energy transfer and the spectra, which yield reliable results as compared to the experiments. However, since the CMRT master equation of an $M$-site system contains a set of $M^2$ ordinary differential equations, the simulation of the dynamics becomes demanding when the system contains hundreds of sites, e.g. 96 sites in PSI. Therefore, we utilize the non-Markovian quantum-jump method \cite{Plenio1998RMP, piilo2008prl, piilo2009pra, Ai2014njp, Breuer2009EPL, Rebentrost2009JCP, Becker2023PRL, Li2025IE} to reduce the computational complexity. The CMRT dynamics is described by the generalized Lindblad term $\mathcal{L}_{\textrm{cmrt}}(\rho)=\sum_{k,p}R_{kp}\mathcal{D}(|\varepsilon_k\rangle\langle \varepsilon_p|)\rho$, with $R_{kp}$ ($k\neq p$) being the energy-transfer rate from the eigen state $|\varepsilon_p\rangle$ to $|\varepsilon_k\rangle$ and $R_{kk}$ being the pure-dephasing rate \cite{Ai2014njp, supp}.

Therefore, on account of the biochemical reaction and the spontaneous emission, the master equation of the CMRT reads
\begin{align}
	\frac{\partial\rho}{\partial t}=&-i[H,\rho]+\mathcal{L}_{\textrm{rc}}(\rho)+\mathcal{L}_{\textrm{em}}(\rho)+\mathcal{L}_{\textrm{cmrt}}(\rho).
\end{align}
The efficiency $\eta$ of the light-harvesting system is \cite{Wu2013PRL}
\begin{align}
	\eta=\Gamma\int_0^{\infty} \langle N|\rho(t)|N\rangle dt. \label{eq3}
\end{align}

To investigate the effect of the dark state in the bridge, we introduce a simplified 3-site bridge model with $N=3$ in the Hamiltonian $H$. The EIT effect exists under the two-photon resonance condition $E_1=E_3$. We use $\Delta=E_2-E_1$ to represent the single-photon detuning between the intermediate site and the other two sites. For a linear-type bridge, the large distance between the initial site 1 and the final site 3 leads to a weak coupling, i.e., $J_{13}\ll J_{12}, J_{23}$. Therefore, we approximately take $J_{13}=0$. By setting $E_1$ as the zero-point of energy, the Hamiltonian reads $H=\Delta|2\rangle\langle 2|+J_{12}|1\rangle\langle 2|+J_{23}|2\rangle\langle 3|+\textrm{h.c.}$ The eigenvalues are $\varepsilon_0=0$ and $\varepsilon_{\pm}=\left(\Delta\pm \lambda\right)/2$, where $J^2=J_{12}^2+J_{23}^2$, $\lambda=\sqrt{\Delta^2+4J^2}$. The eigenvalues can be simplified as $\varepsilon_+=J\tan\phi$ and $\varepsilon_-=-J\cot\phi$ with $\tan2\phi=-2J/\Delta$. By introducing the mixing angle $\theta=\tan^{-1}(J_{12}/J_{23})$, the eigenstates corresponding to $\varepsilon_0$, $\varepsilon_+$, and $\varepsilon_-$ are respectively
$|d\rangle=(\cos\theta,0,-\sin\theta)^T$, 
$|b_+\rangle=(\sin\theta\cos\phi,\sin\phi,\cos\theta\cos\phi)^T$, and
$|b_-\rangle=(\sin\theta\sin\phi,-\cos\phi,\cos\theta\sin\phi)^T$,
where $|d\rangle$ is the dark state and $|b_{\pm}\rangle$ are the two bright states. 

\begin{figure}[!ht]
	\centering
	\includegraphics[width=0.95\linewidth]{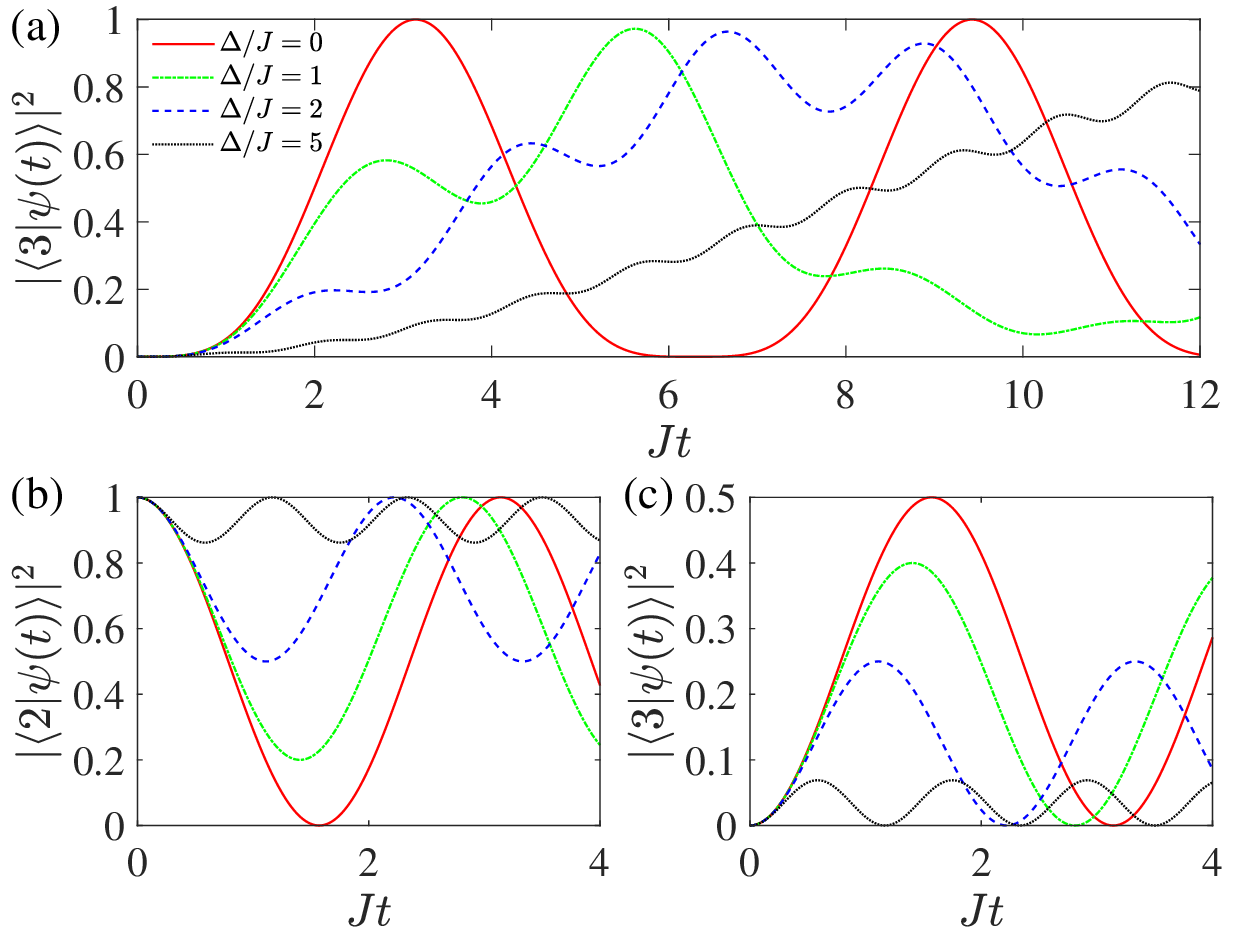}
	\caption{The population dynamics on the intermediate site $|2\rangle$ and the target site $|3\rangle$ with different detunings $\Delta$s, where $J_{12}=J_{23}=J/\sqrt{2}$. The initial state is $|1\rangle$ in (a) and $|2\rangle$ in (b) and (c). Solid red, dash-dotted green, dashed blue, and dotted black lines denote $\Delta/J=0, 1, 2, 5$, respectively. } \label{fig2}
\end{figure}

In this framework, we begin by analytically investigating the dynamics of the bridge while neglecting dissipation. Fig.~\ref{fig2}(a) presents the population transfer from the initial state $|1\rangle$ to the target state $|3\rangle$. For {$|\psi(0)\rangle=|1\rangle=\cos\theta|d\rangle+\sin\theta\cos\phi|b_{+}\rangle+\sin\theta\sin\phi|b_{-}\rangle$}, the population on $|3\rangle$ at time $t$ is
{$|\langle 3|\psi(t)\rangle|^{2}=(\sin^{2}2\theta)/4\cdot|\cos^{2}\phi\exp(-i\varepsilon_{+}t)+\sin^{2}\phi\exp(-i\varepsilon_{-}t)-1|^{2}$}. The fastest transfer occurs at zero single-photon detuning, i.e., $\Delta=0$. The excitation transfer becomes slower with larger $\Delta$. On the other hand, Fig.~\ref{fig2}(b) and (c) highlight the role of the intermediate state $|2\rangle$. When $\Delta=0$, the excitation on $|2\rangle$ can be completely transferred out during the evolution. As $\Delta$ increases, the minimum population on $|2\rangle$ rises and the maximum population on $|3\rangle$ decreases, indicating that more excitation is trapped on site~2 and therefore can not reach site~3. To maximize the energy-transfer efficiency, $\Delta$ should be tuned to zero. Such tuning of the site energy can be implemented via the Stark effect, \cite{Saffman2010RMP, Quirk2024PRL, Empedocles1997Science, LaMountain2021nc, Germanis2016PRAppl, Aghaeimeibodi2019APL, Klein2016NL, Chakraborty2017NL, Roch2018NL, Noh2018NL, Bassett2011PRL}, e.g. by applying a local electric field to the relevant site.

\begin{figure}[!ht]
	\centering
	\includegraphics[bb=0 0 420 315,width=8.5cm]{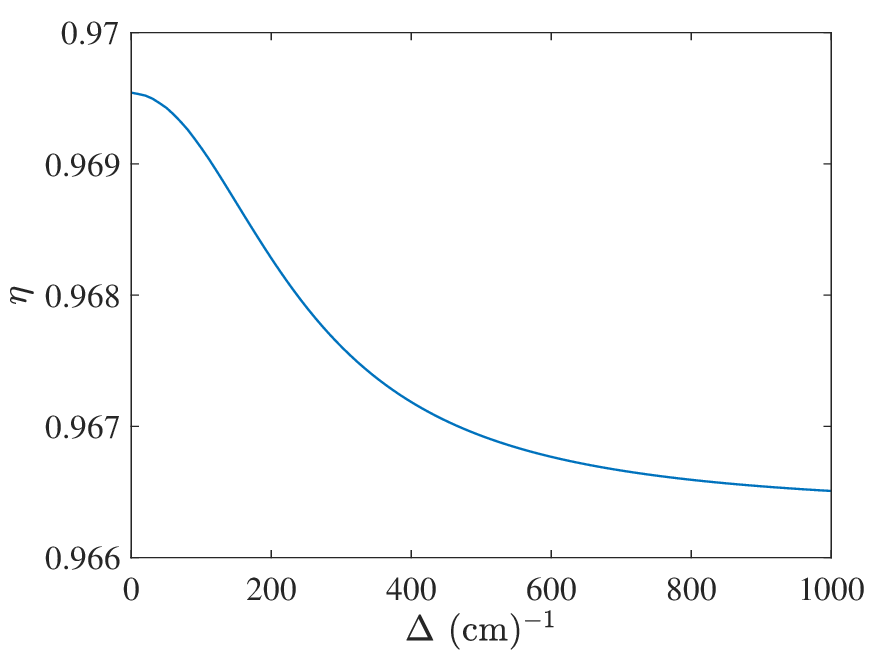}
	\caption{The effect of the detuning of the intermediate site $\Delta$ on the energy transfer efficiency $\eta$, with $J_{12}=55.57~\textrm{cm}^{-1}$, $J_{23}=-37.04~\textrm{cm}^{-1}$, and $J_{13}=-0.36~\textrm{cm}^{-1}\ll J_{12}, ~J_{23}$ for the cluster A4-A28-A40 in PSI, $\kappa^{-1}=1$~ns, $\Gamma'^{-1}=10$~ps. } \label{fig3}
\end{figure}

\begin{figure}[!ht]
	\centering
	\includegraphics[width=\linewidth]{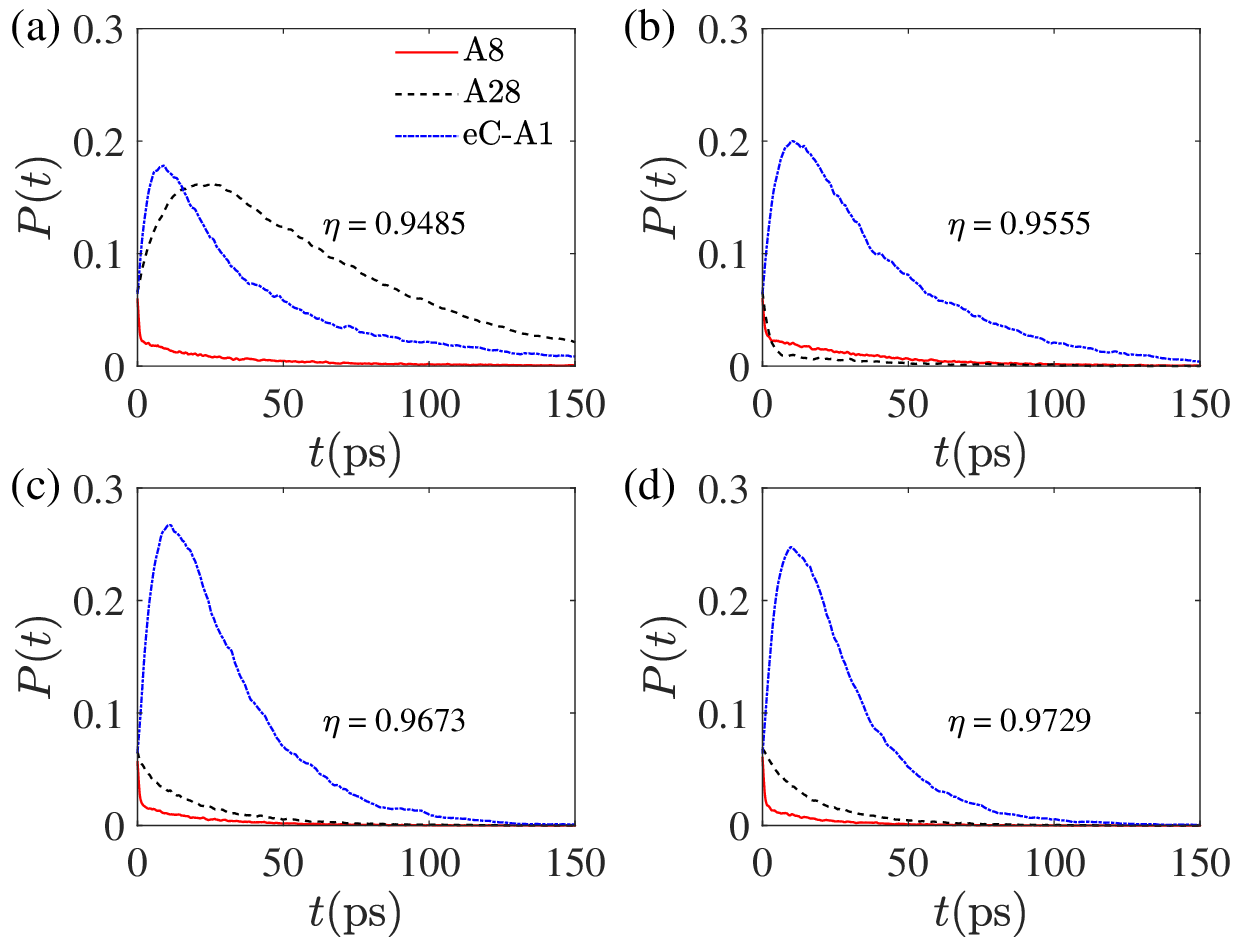}
	\caption{The population dynamics of the OA, the RC and the bridge with $\kappa^{-1}=1$~ns and $\Gamma^{-1}=10$~ps. The level structure is (a) ``extremely bad" with $\Delta=1000~\textrm{cm}^{-1}$, (b) the same as the natural PSI, (c) modified to $E_1=E_2=E_3=14006~\textrm{cm}^{-1}$, (d) {modified to $E_{1}=E_{2}=E_{3}=13940~\textrm{cm}^{-1}$, whose efficiency $\eta=0.9729$ is the highest in the parameter regime.}} \label{fig4}
\end{figure}

\section{Efficiency in the natural PSI}


For a natural photosynthesis system such as PSI which contains 96 sites, the efficiency is optimized during millions of years of evolution. To highlight the effect of our level-modulating proposal, we only investigate part of the total system, as shown in Fig.~\ref{fig:scheme}(b). The partial network contains six sites in the OA, i.e., A8, A10, A11, A18, A19 and A20, the RC with six sites, and the bridge, i.e., A4, A28 and A40, whose energies of the excited state are modified in our scheme while the couplings are the same as the natural PSI. We first investigate the mechanism of the simplified three-site bridge model shown in Fig.~\ref{fig:scheme}(a), where A4, A28, and A40 correspond to sites 1, 2, and 3, respectively. The coupling constants are $J_{12}=43.72~\textrm{cm}^{-1}$, $J_{23}=12.72~\textrm{cm}^{-1}$, and $J_{13}=-0.44~\textrm{cm}^{-1}\ll J_{12},J_{23}$ \cite{Damjanovic2002jpcb}. We tune $E_1=E_3$ to satisfy the two-photon resonance condition for EIT. The spontaneous-emission rate $\kappa^{-1}=1$~ns \cite{Gilmore1998bioch}, and the effective charge-separation rate $\Gamma'^{-1}=10~\textrm{ps}$ \cite{Nelson2015arb}. As shown in Fig.~\ref{fig3}, the efficiency decreases as the detuning $\Delta=E_2-E_1$ increases, because more energy becomes trapped on site 2, enhancing the effect of dissipation.



Next, we connect the OA and the RC through the bridge A4-A28-A40. We investigate the evolution of the system via the CMRT, where the population-transfer rates $R_{ij}$'s are derived from the Hamiltonian of the 15-site network \cite{Hwangfu2015CP}. The spontaneous-emission rate is $\kappa^{-1}=1$~ns \cite{Gilmore1998bioch}. We consider eC-A1 in the RC as the destination of the energy transfer, whose charge-separation rate is $\Gamma=10^{-1}~\textrm{ps}$ \cite{Nelson2015arb}. The efficiency $\eta$ of the light-harvesting network is derived by integrating the population on the excited state of the RC site eC-A1, i.e.,  $\eta=\Gamma\int_0^{\infty} \langle \textrm{eC-A1}|\rho(t)|\textrm{eC-A1}\rangle dt$. The initial state is the maximum mix state of all 15 sites, i.e., $\rho(0)=I/15$. In Fig.~\ref{fig4}, we present the population dynamics on A8 in the OA, the intermediate site A28 in the bridge, and the RC site eC-A1. In Fig.~\ref{fig4}(a), we investigate the large-detuning case with two-photon resonance, i.e., $E_1=E_3$ and $\Delta=1000~\textrm{cm}^{-1}\gg J_{12},~J_{23}$. The efficiency is $\eta=0.9485$. For comparison, we also present the dynamics of the system where the excitation energy of each site is the same as in the natural PSI, as shown in Fig.~\ref{fig4}(b). The efficiency of the large-detuning case is lower than the natural PSI. This is because when the detuning $\Delta$ is large, the population on the intermediate site A28 cannot be effectively transported to other sites, which has been demonstrated in Fig.~\ref{fig2}. Thus, in this condition, A28 becomes an obstacle of the energy-transfer process, hence limits the efficiency of the whole system. 

\begin{figure}[!ht]
	\centering
	\includegraphics[width=\linewidth]{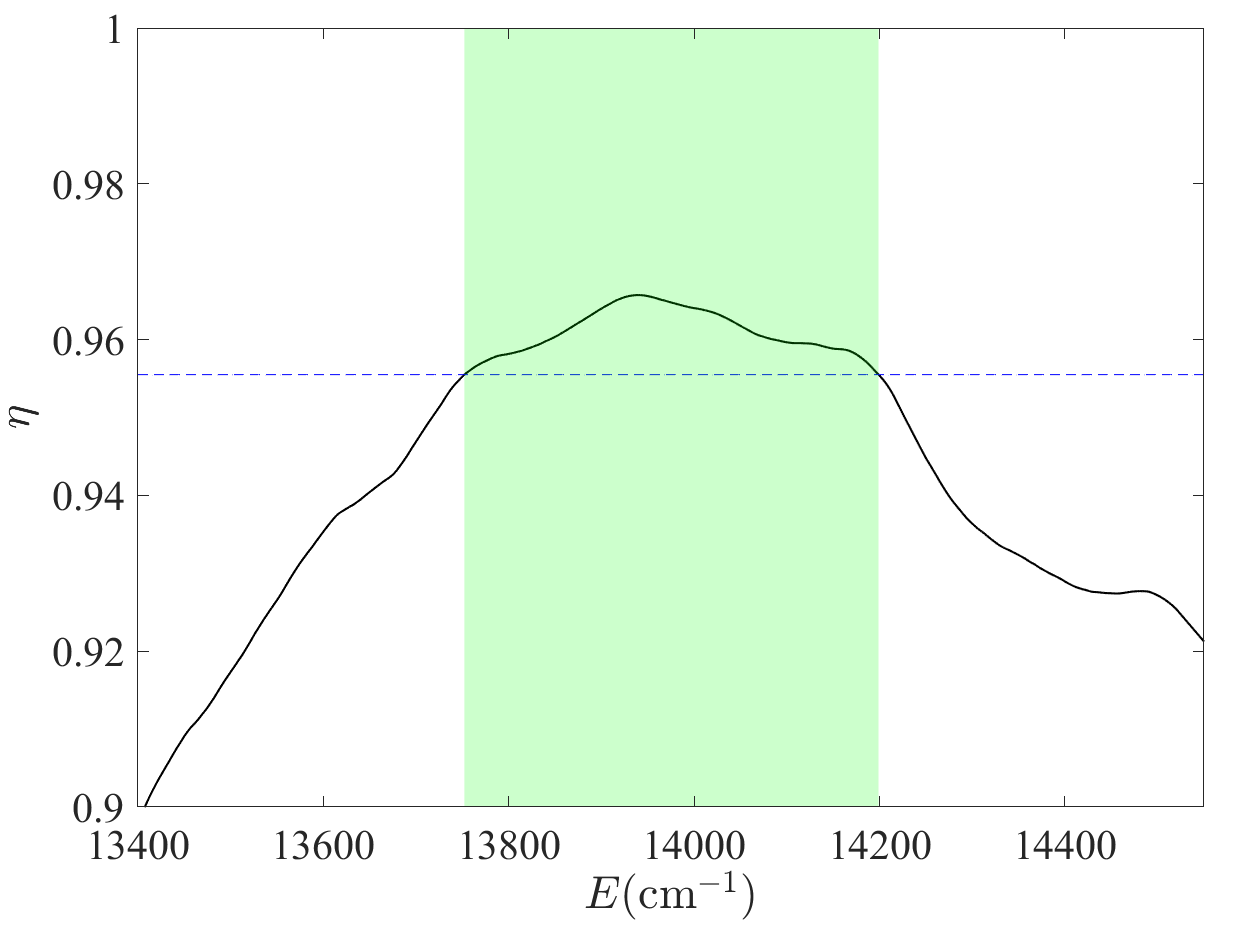}
	\caption{The energy-transfer efficiency of the light-harvesting network where the energy levels of the bridge are modified to be $E_1=E_2=E_3=E$. Dashed line represents the efficiency where the energy levels of the bridge are the same as the natural PSI. In the green area, i.e. {$13752 ~\textrm{cm}^{-1}< E < 14199 ~\textrm{cm}^{-1}$}, the bridge with modified energy levels works better than the natural PSI. } \label{fig5}
\end{figure}

Finally, we modify the level structure of each subsystem to satisfy both single-photon and two-photon resonance, i.e., $E_1=E_2=E_3=E$, where $E=14006~\textrm{cm}^{-1}$ is the site energy of A4 in natural PSI. As shown in Fig.~\ref{fig4}(c), when the exited energies of the three bridge sites are on resonance, the populations on the OA sites and bridge sites decrease more rapidly than the cases in Fig.~\ref{fig4}(a), which indicates that the energy are transferred to the RC more quickly. Therefore, the spontaneous-emission decay of the OA and bridge sites is decreased if the level structure of the 3-site cluster is resonant, and the energy-transfer efficiency of the network increases. The results in Fig.~\ref{fig4} indicate that the energy-transfer efficiency of the whole system will increase if we modulate the level structure of the 3-site cluster to be resonant. In addition, we also investigate the influence of $E$ on the efficiency, as shown in Fig.~\ref{fig5}. When {$13752 ~\textrm{cm}^{-1}< E < 14199 ~\textrm{cm}^{-1}$}, the bridge with modified energy levels works better than the natural PSI. The optimal range lies between the lowest excited-state energy of the RC, i.e., 13201 $\textrm{cm}^{-1}$ and the highest excited-state energy of the OA i.e., 14513 $\textrm{cm}^{-1}$. The maximum efficiency, $\eta=0.9729$, is achieved at $E=13940~\textrm{cm}^{-1}$, as shown in Fig.~\ref{fig4}(d).

\begin{figure}[!ht]
	\centering
	\includegraphics[width=\linewidth]{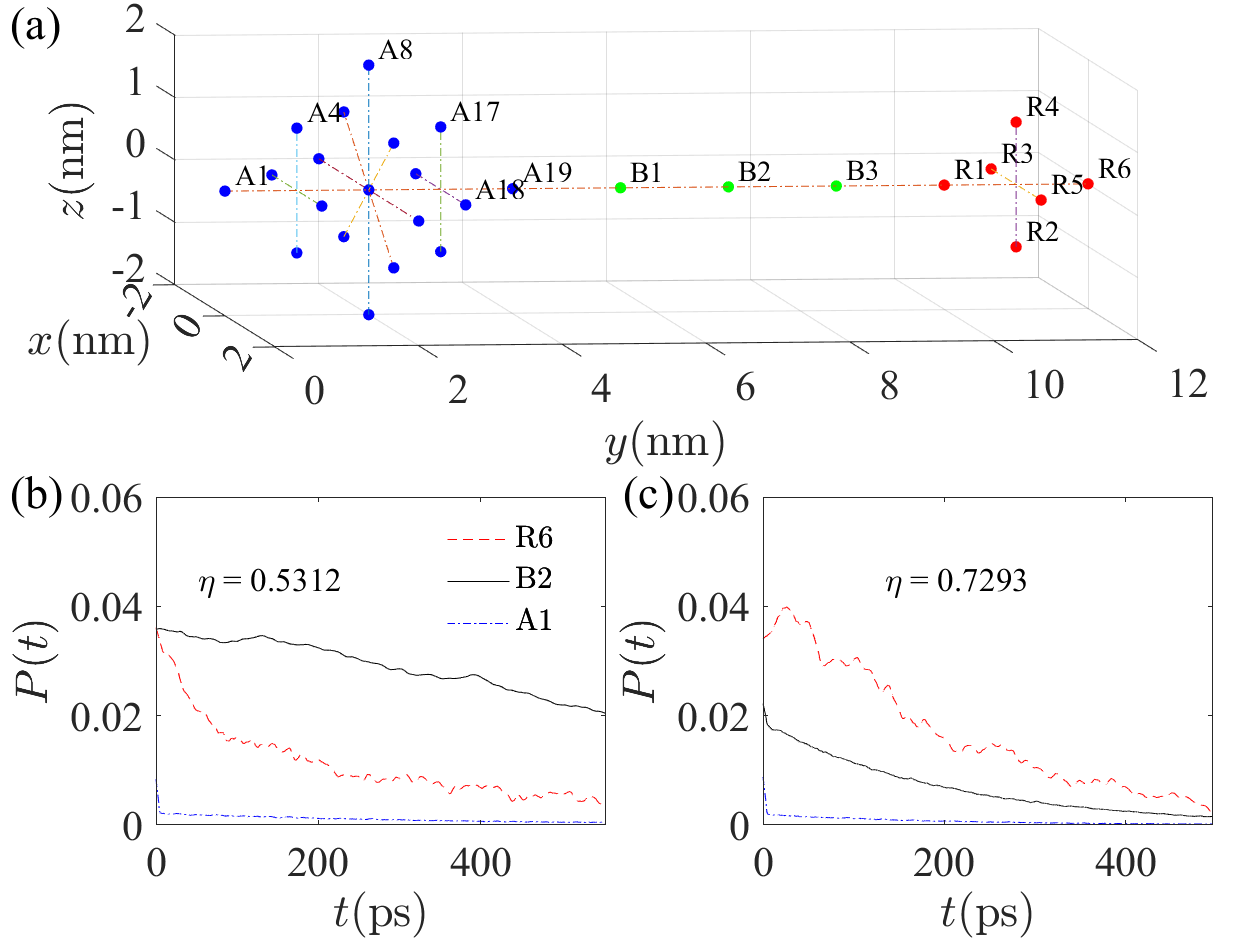}
	\caption{(a) Diagram of the artificial light-harvesting system. The OA contains 19 sites, i.e., A1 to A19. The RC contains 6 sites i.e., R1 to R6. The bridge contains 3 sites i.e., B1 to B3. The dipole moment of each site is 5.2384 Debye. The dipole-moment orientation of A1-A19 and B1 is alone the $x$ direction, while the orientation of R1-R6 and B3 is alone the {$z$ direction}. The dipole-moment orientation of B2 is alone the bisector of the $x$ and $z$ axis. (b)-(c) The population dynamics of the OA, the RC and the bridge with $\kappa^{-1}=1$~ns and $\Gamma^{-1}=10$~ps. The level structure is (b) ``extremely bad" with $\Delta=1000~\textrm{cm}^{-1}$, (c) modified to $E_1=E_2=E_3=14100~\textrm{cm}^{-1}$. } \label{fig6}
\end{figure}

\section{Efficiency in the artificial light-harvesting system}


Furthermore, we investigate the performance of the bridge in an artificial light-harvesting system. The OA contains 19 sites that form a regular octahedron with side length being $2\sqrt{2}$~nm. The RC contains 6 sites that also form a regular octahedron with side length equals to $\sqrt{2}$~nm. The bridge contains 3 sites. The distance between the nearest neighbor in the A19-B1-B2-B3-R1 chain is 1.5~nm. The dipole moment of each site is 5.2384~Debye \cite{Damjanovic2002jpcb}. The dipole-moment orientation of A1-A19 and B1 is alone the $x$ direction, while the orientation of R1-R6 and B3 is alone the {$z$ direction}. The dipole-moment orientation of B2 is alone the bisector of the $x$ and $z$ axis. We assume the charge-separation process occurs on R6.


We first insert a large-detuning bridge with site energies being respectively $E_1=E_3=14100~\textrm{cm}^{-1}$ and $E_2=13100~\textrm{cm}^{-1}$. Here $E_1$, $E_2$ and $E_3$ are the site energy of B1, B2 and B3, respectively. The initial state is the maximum mixed state of all 28 sites, i.e., $\rho(0)=I/28$. The efficiency is 0.5312. The population on B2, i.e., the intermediate site of the bridge, can not be efficiently transferred to the other sites, as shown in Fig.~\ref{fig6}(b). Next, we modify the site energy $E_2$ to be resonant with $E_1$ and $E_3$. In this case, the populations on the OA and the bridge are transferred to the RC much faster than the large-detuning case, as shown in Fig.~\ref{fig6}(c). Therefore, the efficiency is increased to 0.7293.

\section{Conclusion and remarks}


In this paper, we investigate the energy-transfer process in both natural and artificial photosynthetic systems. We focus on the bridge connecting the RC and the OA far away from the RC. We find that by modulating the level structure of the 3-site bridge to be resonant, the energy transfer will become more quickly and efficiently with the assistance of the dark state, and the total energy-transfer efficiency will be enhanced.

\section{Acknowledgement}

This work is supported by the Quantum Science and Technology-National Science and Technology Major Project (2023ZD0300200), and National Natural Science Foundation of China under Grant No.~62461160263, and Guangdong Provincial Quantum Science Strategic Initiative under Grant No.~GDZX2505004.
	
%
	
	
%

\clearpage
\section{}


\section*{Supplementary material}

# Supplementary Material for "Electromagnetically-Induced Transparency Bridges Disconnected Light-Harvesting Network"

Jun Wang,[1,2] Rui Li,[1,2] Yi Li,[1,2] Kai-Ya Zhang,[1,2] and Qing Ai[1,2,*]

[1]*School of Physics and Astronomy, Applied Optics Beijing Area Major Laboratory,*
*Beijing Normal University, Beijing 100875, China*
[2]*Key Laboratory of Multiscale Spin Physics, Ministry of Education,*
*Beijing Normal University, Beijing 100875, China*

## I. THE NON-HERMITIAN HAMILTONIAN METHOD

Instead of the master equation in the main text, here we use the non-Hermitian Hamiltonian method to calculate the energy transfer efficiency $\eta$ for the two cases with $\Delta = 0$ and $\Delta \gg J$. First of all, we obtain the analytical solution under some approximation, and then compare the analytical result with the one by the numerically-exact calculation to demonstrate the reasonability of our analytical result.

### A. The resonant case with $\Delta = 0$

When the energy gap $\Delta$ vanishes, the non-Hermitian Hamiltonian of the cluster reads ($\hbar = 1$)

$$H = \begin{pmatrix} -i\kappa & J_{12} & J_{13} \\ J_{12} & -i\kappa & J_{23} \\ J_{13} & J_{23} & -i(\kappa + \Gamma) \end{pmatrix}, \tag{S1}$$

where $\kappa$ and $\Gamma$ are respectively the dissipation rate and the chemical-reaction rate, satisfying $\kappa \ll \Gamma \ll J_{12}, J_{23}, J_{13}$. The non-Hermitian parts describe the spontaneous emission and the chemical reaction. Noted that

$$H = H_0 - i\kappa I, \tag{S2}$$

where $I$ is the unitary matrix and Hamiltonian $H'$ is

$$H_0 = \begin{pmatrix} 0 & J_{12} & J_{13} \\ J_{12} & 0 & J_{23} \\ J_{13} & J_{23} & -i\Gamma \end{pmatrix}. \tag{S3}$$

---

* aiqing@bnu.edu.cn

Therefore, the eigenvalue of Hamiltonian $H$ can be written as

$$\lambda = x - i\kappa, \tag{S4}$$

where $x$ is the eigenvalue of Hamiltonian $H_0$. The eigen function of $x$ is

$$x^3 + i\Gamma x^2 - (J_{12}^2 + J_{23}^2 + J_{13}^2)x - iJ_{12}^2\Gamma - 2J_{12}J_{23}J_{13} = 0 \tag{S5}$$

Next we consider the situations in the main text. The coupling coefficients $J_{12}, J_{23}, J_{13}$ satisfy

$$J_{12} \sim J_{23} \gg J_{13}. \tag{S6}$$

The energy-level diagram of the system is shown in Fig. S1.

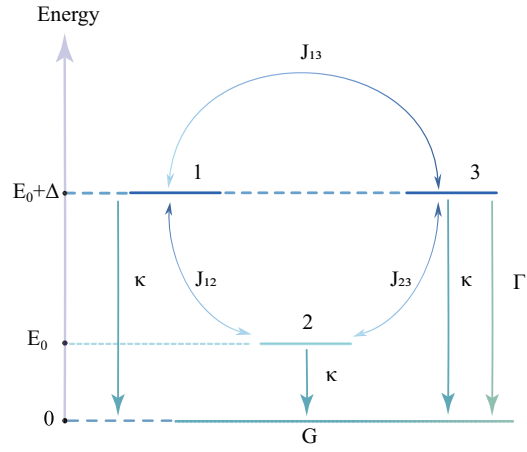

FIG. S1. Energy-level diagram of the cluster, with the energy gap $\Delta$, and couplings $J_{12}, J_{13}$ and $J_{23}$. The energy of ground state $|G\rangle$ is set to zero.

In this situation, the parameters satisfy $\Gamma \sim J_{13} \ll J_{12}, J_{23}$. Thus, Eq. (S5) becomes

$$x^3 + i\Gamma x^2 - J^2 x - iJ_{12}^2 \Gamma - 2J_{12}J_{23}J_{13} = 0, \tag{S7}$$

where $J^2 = J_{12}^2 + J_{23}^2$. Since this equation contains both real and imaginary parts, we divide $x$ into the real and imaginary parts as

$$x = p + qi. \tag{S8}$$

Substituting Eq. (S8) into (S7), we have

$$p^3 - 3pq^2 - 2pq\Gamma - J^2 p - 2J_{12}J_{23}J_{13} = 0, \tag{S9}$$

$$q^3 - 3p^2 q - (p^2 - q^2)\Gamma + J^2 q + J_{12}^2 \Gamma = 0. \tag{S10}$$

Equation (S9) can be rewritten as

$$p(p^2 - 3q^2 - 2q\Gamma - J^2) = 2J_{12}J_{23}J_{13}. \tag{S11}$$

The right hand side of Eq. (S11) is of first order. Hereafter, we utilize the perturbation theory [1] to find the approximated solution to this equation, i.e.,

$$p = p^{(0)} + p^{(1)}. \tag{S12}$$

Equation (S11) yields two equations, i.e., the zero-order equation and the first-order equation,

$$p^{(0)}[(p^{(0)})^2 - 3q^2 - 2q\Gamma - J^2] = 0, \tag{S13}$$

$$p^{(1)}[(p^{(0)})^2 - 3q^2 - 2q\Gamma - J^2] + 2(p^{(0)})^2 p^{(1)} = 2J_{12}J_{23}J_{13}. \tag{S14}$$

The solutions of Eq. (S13) are

$$p_1^{(0)} = 0, \tag{S15}$$

$$p_{2,3}^{(0)} = \pm\sqrt{3q^2 + 2q\Gamma + J^2}. \tag{S16}$$

Substituting Eq. (S15) into Eq. (S14), we have

$$p_1^{(1)} = \frac{-2J_{12}J_{23}J_{13}}{3q^2 + 2q\Gamma + J^2}, \tag{S17}$$

$$p_2^{(1)} = p_3^{(1)} = \frac{J_{12}J_{23}J_{13}}{3q^2 + 2q\Gamma + J^2}. \tag{S18}$$

From Eqs. (S16) and (S17), we find that $p^{(1)}$ is of the same order as $J_{13}$. To the first order of $\Gamma$ and $J_3$, we have

$$p_1 = \frac{-2J_{12}J_{23}J_{13}}{3q^2 + 2q\Gamma + J^2}, \tag{S19}$$

$$p_{2,3} = \pm\sqrt{3q^2 + 2q\Gamma + J^2}. \tag{S20}$$

Substituting these solutions to Eq. (S10), we can obtain

$$-q_1^3 + 3\left(\frac{-2J_{12}J_{23}J_{13}}{3q_1^2 + 2q_1\Gamma + J^2}\right)^2 q_1 + \left[\left(\frac{-2J_{12}J_{23}J_{13}}{3q_1^2 + 2q_1\Gamma + J^2}\right)^2 - q_1^2\right]\Gamma - J^2 q_1 - J_{12}^2\Gamma = 0, \tag{S21}$$

$$-q_{2,3}^3 + 3(3q_{2,3}^2 + 2q_{2,3}\Gamma + J^2)q_{2,3} + (2q_{2,3}^2 + 2q_{2,3}\Gamma + J^2)\Gamma - J^2 q_{2,3} - J_{12}^2\Gamma = 0. \tag{S22}$$

If we assume $q \sim J$, to the zeroth order of $\Gamma$ and $J_3$, the equations yield

$$q_1^2 + J^2 = 0, \tag{S23}$$

$$4q_{2,3}^2 + J^2 = 0, \tag{S24}$$

which is conflict with our assumption that $p$ and $q$ are all real. Thus, $q$ is of the same order as $J_{13}$ and $\Gamma$. To the first order of $\Gamma$ and $J_3$, Eqs. (S20) and (S21) become

$$q_1 = -\frac{J_{12}^2}{J^2}\Gamma, \tag{S25}$$

$$q_2 = q_3 = -\frac{J_{23}^2}{2J^2}\Gamma. \tag{S26}$$

In all, we obtain the three eigenvalues as

$$x_1 = \frac{-2J_{12}J_{23}}{J^2}J_{13} - i\frac{J_{12}^2}{J^2}\Gamma, \tag{S27}$$

$$x_2 = J - i\frac{J_{23}^2}{2J^2}\Gamma, \tag{S28}$$

$$x_3 = -J - i\frac{J_{23}^2}{2J^2}\Gamma. \tag{S29}$$

The eigenvalues of the Hamiltonian $H$ are

$$\lambda_1 = x_1 - i\kappa = \frac{-2J_{12}J_{23}}{J^2}J_{13} - i\left(\frac{J_{12}^2}{J^2}\Gamma + \kappa\right), \tag{S30}$$

$$\lambda_2 = x_2 - i\kappa = J - i\left(\frac{J_{23}^2}{2J^2}\Gamma + \kappa\right), \tag{S31}$$

$$\lambda_3 = x_3 - i\kappa = -J - i\left(\frac{J_{23}^2}{2J^2}\Gamma + \kappa\right). \tag{S32}$$

To the zeroth order of $\Gamma$ and $J_3$, the eigenvectors corresponding to the above eigenvalues are respectively

$$|d\rangle = (\frac{J_{23}}{J}, 0, -\frac{J_{12}}{J})^T, \tag{S33}$$

$$|b_+\rangle = \frac{1}{\sqrt{2}}(\frac{J_{12}}{J}, 1, \frac{J_{23}}{J})^T, \tag{S34}$$

$$|b_-\rangle = \frac{1}{\sqrt{2}}(-\frac{J_{12}}{J}, 1, -\frac{J_{23}}{J})^T, \tag{S35}$$

where $|d\rangle$ is the dark state, while $|b_\pm\rangle$ are the two bright states. Alternatively, the bases $|j\rangle$ ($j = 1, 2, 3$) can be written in terms of the eigenvectors as

$$|1\rangle = \frac{J_{23}}{J}|d\rangle + \frac{\sqrt{2}J_{12}}{2J}(|b_+\rangle - |b_-\rangle), \tag{S36}$$

$$|2\rangle = \frac{1}{\sqrt{2}}(|b_+\rangle + |b_-\rangle), \tag{S37}$$

$$|3\rangle = -\frac{J_{12}}{J}|d\rangle + \frac{\sqrt{2}J_{23}}{2J}(|b_+\rangle - |b_-\rangle). \tag{S38}$$

As a result, when the initial state is $|j\rangle$ ($j = 1, 2, 3$), respectively, the state at time $t$ is

$$|\psi_1(t)\rangle = \frac{J_{23}}{J}|d\rangle e^{-i\lambda_1 t} + \frac{\sqrt{2}J_{12}}{2J}(|b_+\rangle e^{-i\lambda_2 t} - |b_-\rangle e^{-i\lambda_3 t})$$

$$= \frac{J_{23}}{J}|d\rangle \exp\left(i\frac{2J_{12}J_{23}}{J^2}J_{13}t\right)\exp\left[-\left(\frac{J_{12}^2}{J^2}\Gamma + \kappa\right)t\right]$$

$$+ \frac{\sqrt{2}J_{12}}{2J}\left\{|b_+\rangle\exp(-iJt)\exp\left[-\left(\frac{J_{23}^2}{2J^2}\Gamma + \kappa\right)t\right] - |b_-\rangle\exp(iJt)\exp\left[-\left(\frac{J_{23}^2}{2J^2}\Gamma + \kappa\right)t\right]\right\},$$

(S39)

$$|\psi_2(t)\rangle = \frac{1}{\sqrt{2}}(|b_+\rangle e^{-i\lambda_2 t} + |b_-\rangle e^{-i\lambda_3 t})$$

$$= \frac{1}{\sqrt{2}}\left\{|b_+\rangle\exp(-iJt)\exp\left[-\left(\frac{J_{23}^2}{2J^2}\Gamma + \kappa\right)t\right] + |b_-\rangle\exp(iJt)\exp\left[-\left(\frac{J_{23}^2}{2J^2}\Gamma + \kappa\right)t\right]\right\},$$

(S40)

$$|\psi_3(t)\rangle = -\frac{J_{12}}{J}|d\rangle e^{-i\lambda_1 t} + \frac{\sqrt{2}J_{23}}{2J}(|b_+\rangle e^{-i\lambda_2 t} - |b_-\rangle e^{-ix\lambda_3 t})$$

$$= -\frac{J_{12}}{J}|d\rangle \exp\left(i\frac{2J_{12}J_{23}}{J^2}J_{13}t\right)\exp\left[-\left(\frac{J_{12}^2}{J^2}\Gamma + \kappa\right)t\right]$$

$$+ \frac{\sqrt{2}J_{23}}{2J}\left\{|b_+\rangle\exp(-iJt)\exp\left[-\left(\frac{J_{23}^2}{2J^2}\Gamma + \kappa\right)t\right] - |b_-\rangle\exp(iJt)\exp\left[-\left(\frac{J_{23}^2}{2J^2}\Gamma + \kappa\right)t\right]\right\}.$$

(S41)

When the probabilities of the initial states being $|j\rangle$ ($j = 1, 2, 3$) are equal, the population on $|3\rangle$ at time $t$ is

$$P_3(t) = \frac{1}{3}\left[|\langle 3|\psi_1(t)\rangle|^2 + |\langle 3|\psi_2(t)\rangle|^2 + |\langle 3|\psi_3(t)\rangle|^2\right]$$

$$= \frac{1}{3}\left\{\left(\frac{J_{12}J_{23}}{J^2}\right)^2\left[\exp\left[-\left(\frac{2J_{12}^2}{J^2}\Gamma + \kappa\right)t\right] + \cos^2 Jt\exp\left[-\left(\frac{J_{23}^2}{J^2}\Gamma + \kappa\right)t\right]\right]\right.$$

$$+ \frac{J_{23}^2}{J^2}\sin^2 Jt\exp\left[-\left(\frac{J_{23}^2}{J^2}\Gamma + \kappa\right)t\right]$$

$$\left.+ \left(\frac{J_{12}}{J}\right)^4\exp\left[-\left(\frac{2J_{12}^2}{J^2}\Gamma + \kappa\right)t\right] + \left(\frac{J_{23}}{J}\right)^4\cos^2 Jt\exp\left[-\left(\frac{J_{23}^2}{J^2}\Gamma + \kappa\right)t\right]\right\}$$

$$= \frac{1}{3}\left\{\frac{J_{12}^2}{J^2}\exp\left[-\left(\frac{2J_{12}^2}{J^2}\Gamma + \kappa\right)t\right] + \frac{J_{23}^2}{J^2}\exp\left[-\left(\frac{J_{23}^2}{J^2}\Gamma + \kappa\right)t\right]\right\}.$$

(S42)

The efficiency $\eta$ of this 3-level-system is

$$\eta = 2\Gamma\int_0^\infty \langle 3|\rho(t)|3\rangle dt$$

$$= 2\Gamma\int_0^\infty P_3(t)dt,$$

$$= \frac{2\Gamma}{3}\left(\frac{1}{2\Gamma + \frac{J^2}{J_{12}^2}\kappa} + \frac{1}{\Gamma + \frac{J^2}{J_{23}^2}\kappa}\right).$$

(S43)

Comparing to the corresponding formula in the main text, Eq. (S43) has an additional coefficient 2 [2]. The reason for this inconsistency is the definition of the chemical-reaction rate $\Gamma$ in master equation and in our non-Hermitian Hamiltonian are not the same. Therefore we compensate for the impact of the difference through adding a coefficient 2 in the definition of $\eta$. We set the values of the parameters as $J_{12} = 132.9$ cm$^{-1}$, $J_{23} = 82.28$ cm$^{-1}$, $J_{13} = 1.907$ cm$^{-1}$, $\Gamma = 0.1$ ps$^{-1} = 3.33$ cm$^{-1}$, $\kappa = 1 \times 10^{-3}$ ps$^{-1} = 3.33 \times 10^{-2}$ cm$^{-1}$ [3]. By numerical simulations, we have

$$\eta = 0.974. \tag{S44}$$

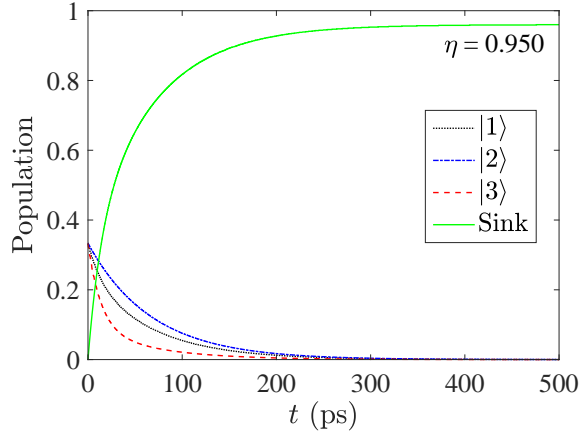

FIG. S2. The population dynamics of $|j\rangle$ ($j = 1, 2, 3$) at time $t$ and the efficiency $\eta = 0.950$ from numerical calculation, with $\Delta = 0$, $J_{12} = 132.9$ cm$^{-1}$, $J_{23} = 82.28$ cm$^{-1}$, $J_{13} = 1.907$ cm$^{-1}$, $\Gamma = 0.1$ ps$^{-1} = 3.33$ cm$^{-1}$, $\kappa = 1 \times 10^{-3}$ ps$^{-1} = 3.33 \times 10^{-2}$ cm$^{-1}$.

As shown, the two results of the efficiency $\eta$ fit quite well. The evolution duration is around 10 ps, on the same scale as $\Gamma^{-1}$.

**B. The large-detuning case with $\Delta \gg J$**

When the energy gap $\Delta$ is much larger compared to $J_{12}, J_{23}, J_{13}$, the non-Hermitian Hamiltonian of the cluster reads

$$H = \begin{pmatrix} -i\kappa & J_{12} & J_{13} \\ J_{12} & \Delta - i\kappa & J_{23} \\ J_{13} & J_{23} & -i(\kappa + \Gamma) \end{pmatrix} = H' - i\kappa I, \tag{S45}$$

where the Hamiltonian $H'$ is

$$H' = \begin{pmatrix} 0 & J_{12} & J_{13} \\ J_{12} & \Delta & J_{23} \\ J_{13} & J_{23} & -i\Gamma \end{pmatrix}, \tag{S46}$$

satisfying $\Gamma \ll J_{12}, J_{23}, J_{13} \ll \Delta$. The Schrödinger equation of the system is

$$H|\psi\rangle = E|\psi\rangle = (E' - i\kappa)|\psi\rangle, \tag{S47}$$

where $E$ is the eigenvalue of $H$ and $E'$ is the eigenvalue of $H'$. Alternatively, we could write

$$H'|\psi\rangle = E'|\psi\rangle, \tag{S48}$$

$$E = E' - i\kappa. \tag{S49}$$

Next we consider the situations in the main text. The coupling coefficients $J_{12}, J_{23}, J_{13}$ satisfy

$$J_{12} \sim J_{23} \gg J_{13}. \tag{S50}$$

In this situation, the parameters satisfy $\kappa \ll \Gamma \sim J_{13} \ll J_{12} \sim J_{23} \ll \Delta$. In order to use the perturbation theory, we divide $H'$ into two parts as

$$H_0 = \begin{pmatrix} 0 & J_{12} & 0 \\ J_{12} & \Delta & J_{23} \\ 0 & J_{23} & 0 \end{pmatrix}, \tag{S51}$$

$$H_1 = \begin{pmatrix} 0 & 0 & J_{13} \\ 0 & 0 & 0 \\ J_{13} & 0 & -i\Gamma \end{pmatrix}. \tag{S52}$$

Therefore, $H_1$ can be regarded as a perturbation. We rewrite the Schrödinger equation as

$$(H_0 + H_1)(|\psi^{(0)}\rangle + |\psi^{(1)}\rangle) = (E^{(0)} + E^{(1)})(|\psi^{(0)}\rangle + |\psi^{(1)}\rangle), \tag{S53}$$

where $\psi^{(0)}$ and $\psi^{(1)}$ are the zeroth-order and the first-order term of $\psi$, $E^{(0)}$ and $E^{(1)}$ are the zeroth-order and the first-order term of $E'$. Expanding Eq. (S53), we obtain the zeroth-order and the first-order equation as

$$H_0|\psi^{(0)}\rangle = E^{(0)}|\psi^{(0)}\rangle, \tag{S54}$$

$$H_0|\psi^{(1)}\rangle + H_1|\psi^{(0)}\rangle = E^{(0)}|\psi^{(1)}\rangle + E^{(1)}|\psi^{(0)}\rangle. \tag{S55}$$

$E^{(0)}$ is determined by

$$E^{(0)}((E^{(0)})^2 - \Delta E^{(0)} - J^2) = 0. \tag{S56}$$

The three eigenvalues are respectively

$$E_1^{(0)} = 0, \tag{S57}$$

$$E_2^{(0)} = \frac{1}{2}(\Delta + \sqrt{\Delta^2 + 4J^2}), \tag{S58}$$

$$E_3^{(0)} = -\frac{1}{2}(\sqrt{\Delta^2 + 4J^2} - \Delta). \tag{S59}$$

The three eigenvectors are correspondingly

$$|\psi_1^{(0)}\rangle = \left(-\frac{J_{23}}{J}, 0, \frac{J_{12}}{J}\right)^T, \tag{S60}$$

$$|\psi_2^{(0)}\rangle = \frac{1}{\sqrt{2(\Delta^2 + 4J^2 + \Delta\sqrt{\Delta^2 + 4J^2})}}(2J_{12}, \Delta + \sqrt{\Delta^2 + 4J^2}, 2J_{23})^T, \tag{S61}$$

$$|\psi_3^{(0)}\rangle = \frac{1}{\sqrt{2(\Delta^2 + 4J^2 - \Delta\sqrt{\Delta^2 + 4J^2})}}(2J_{23}, -(\sqrt{\Delta^2 + 4J^2} - \Delta), 2J_{23})^T. \tag{S62}$$

Next we consider the solution of Eq. (S55). Assuming that

$$|\psi_n^{(1)}\rangle = \sum_{m\neq n} c_m |\psi_m^{(0)}\rangle, \tag{S63}$$

Equation (S55) becomes

$$H_0 \sum_{m\neq n} c_m |\psi_m^{(0)}\rangle + H_1 |\psi_n^{(0)}\rangle = E_n^{(0)} \sum_{m\neq n} c_m |\psi_m^{(0)}\rangle + E_n^{(1)} |\psi_n^{(0)}\rangle,$$

$$\sum_{m\neq n} c_m E_m^{(0)} |\psi_m^{(0)}\rangle + H_1 |\psi_n^{(0)}\rangle = E_n^{(0)} \sum_{m\neq n} c_m |\psi_m^{(0)}\rangle + E_n^{(1)} |\psi_n^{(0)}\rangle. \tag{S64}$$

Multiplying Eq. (S64) by $\langle\psi_n^{(0)}|$ from the left hand side, we have

$$E_n^{(1)} = \langle\psi_n^{(0)}|H_1|\psi_n^{(0)}\rangle. \tag{S65}$$

After some algebra, we obtain

$$E_1^{(1)} = \langle\psi_1^{(0)}|H_1|\psi_1^{(0)}\rangle = -\frac{2J_{12}J_{23}}{J^2}J_{13} - i\frac{J_{12}^2}{J^2}\Gamma, \tag{S66}$$

$$E_2^{(1)} = \langle\psi_2^{(0)}|H_1|\psi_2^{(0)}\rangle = \frac{4J_{12}J_{23}J_{13} - 2iJ_{23}^2\Gamma}{\Delta^2 + 4J^2 + \Delta\sqrt{\Delta^2 + 4J^2}}, \tag{S67}$$

$$E_3^{(1)} = \langle\psi_3^{(0)}|H_1|\psi_3^{(0)}\rangle = \frac{4J_{12}J_{23}J_{13} - 2iJ_{23}^2\Gamma}{\Delta^2 + 4J^2 - \Delta\sqrt{\Delta^2 + 4J^2}}. \tag{S68}$$

To the lowest-order term, the eigenvalues are

$$E_1 = -i\left(\frac{J_{12}^2}{J^2}\Gamma + \kappa\right), \tag{S69}$$

$$E_2 = \Delta - i\left(\frac{J_{23}^2}{\Delta^2}\Gamma + \kappa\right), \tag{S70}$$

$$E_3 = -i\left(\frac{J_{23}^2}{J^2}\Gamma + \kappa\right). \tag{S71}$$

The eigenvectors are

$$|\psi_1\rangle = (-\frac{J_{23}}{J}, 0, \frac{J_{12}}{J})^T, \tag{S72}$$

$$|\psi_2\rangle = \frac{1}{\sqrt{J^2 + \Delta^2}}(J_{12}, \Delta, J_{23})^T, \tag{S73}$$

$$|\psi_3\rangle = \frac{1}{\sqrt{J^2 + \Delta^2}}(J_{12}, -\frac{J^2}{\Delta}, J_{23})^T. \tag{S74}$$

Obviously, the three eigenvectors are orthogonal to each other, i.e.,

$$\langle\psi_1|\psi_2\rangle = \frac{1}{\sqrt{J^2+\Delta^2}}(-\frac{J_{12}J_{23}}{J}+0+\frac{J_{12}J_{23}}{J}) = 0, \tag{S75}$$

$$\langle\psi_1|\psi_3\rangle = \frac{1}{\sqrt{J^2+\Delta^2}}(-\frac{J_{12}J_{23}}{J}+0+\frac{J_{12}J_{23}}{J}) = 0, \tag{S76}$$

$$\langle\psi_2|\psi_3\rangle = \frac{1}{J^2+\Delta^2}(J_{12}^2 - J^2 + J_{23}^2) = 0. \tag{S77}$$

The bases $\{|j\rangle \; j=1,2,3\}$ can be expanded by $\{|\psi_j\rangle, j=1,2,3\}$ as

$$|1\rangle = -\frac{J_{23}}{J}|\psi_1\rangle + \frac{J_{12}}{\sqrt{J^2+\Delta^2}}|\psi_2\rangle + \frac{J_{12}\Delta^2}{J^2\sqrt{J^2+\Delta^2}}|\psi_3\rangle, \tag{S78}$$

$$|2\rangle = \frac{\Delta}{\sqrt{J^2+\Delta^2}}(|\psi_2\rangle - |\psi_3\rangle), \tag{S79}$$

$$|3\rangle = \frac{J_{12}}{J}|\psi_1\rangle + \frac{J_{23}}{\sqrt{J^2+\Delta^2}}|\psi_2\rangle + \frac{J_{23}\Delta^2}{J^2\sqrt{J^2+\Delta^2}}|\psi_3\rangle. \tag{S80}$$

As a result, when the initial state is $|j\rangle$ ($j=1,2,3$), respectively, the state $|\varphi_j(t)\rangle$ ($j=1,2,3$) at time $t$ are

$$\begin{aligned}|\varphi_1(t)\rangle &= -\frac{J_{23}}{J}|\psi_1\rangle e^{-iE_1 t} + \frac{J_{12}}{\sqrt{J^2+\Delta^2}}|\psi_2\rangle e^{-iE_2 t} + \frac{J_{12}\Delta^2}{J^2\sqrt{J^2+\Delta^2}}|\psi_3\rangle e^{-iE_3 t},\\
&= -\frac{J_{23}}{J}|\psi_1\rangle \exp\left[-\left(\frac{J_{12}^2}{J^2}\Gamma+\kappa\right)t\right] + \frac{J_{12}}{\sqrt{J^2+\Delta^2}}|\psi_2\rangle \exp(-i\Delta t)\exp\left[-\left(\frac{J_{23}^2}{\Delta^2}\Gamma+\kappa\right)t\right]\\
&\quad + \frac{J_{12}\Delta^2}{J^2\sqrt{J^2+\Delta^2}}|\psi_3\rangle \exp\left[-\left(\frac{J_{23}^2}{J^2}\Gamma+\kappa\right)t\right],\end{aligned} \tag{S81}$$

$$\begin{aligned}|\varphi_2(t)\rangle &= \frac{\Delta}{\sqrt{J^2+\Delta^2}}(|\psi_2\rangle e^{-iE_2 t} - |\psi_3\rangle e^{-iE_3 t}),\\
&= \frac{\Delta}{\sqrt{J^2+\Delta^2}}\left\{|\psi_2\rangle \exp(-i\Delta t)\exp\left[-\left(\frac{J_{23}^2}{\Delta^2}\Gamma+\kappa\right)t\right] - |\psi_3\rangle \exp\left[-\left(\frac{J_{23}^2}{J^2}\Gamma+\kappa\right)t\right]\right\},\end{aligned} \tag{S82}$$

$$\begin{aligned}|\varphi_3(t)\rangle &= \frac{J_{12}}{J}|\psi_1\rangle e^{-iE_1 t} + \frac{J_{23}}{\sqrt{J^2+\Delta^2}}|\psi_2\rangle e^{-iE_2 t} + \frac{J_{23}\Delta^2}{J^2\sqrt{J^2+\Delta^2}}|\psi_3\rangle e^{-iE_3 t},\\
&= \frac{J_{12}}{J}|\psi_1\rangle \exp\left[-\left(\frac{J_{12}^2}{J^2}\Gamma+\kappa\right)t\right] + \frac{J_{23}}{\sqrt{J^2+\Delta^2}}|\psi_2\rangle \exp(-i\Delta t)\exp\left[\left(\frac{J_{23}^2}{\Delta^2}\Gamma+\kappa\right)t\right]\\
&\quad + \frac{J_{23}\Delta^2}{J^2\sqrt{J^2+\Delta^2}}|\psi_3\rangle \exp\left[-\left(\frac{J_{23}^2}{J^2}\Gamma+\kappa\right)t\right].\end{aligned} \tag{S83}$$

The probability amplitudes of $|\varphi_j(t)\rangle$ ($j=1,2,3$) at $|3\rangle$ are respectively

$$\begin{aligned}\langle 3|\varphi_1(t)\rangle =& -\frac{J_{23}}{J}\langle 3|\psi_1\rangle \exp\left[-\left(\frac{J_{12}^2}{J^2}\Gamma+\kappa\right)t\right]+\frac{J_{12}}{\sqrt{J^2+\Delta^2}}\langle 3|\psi_2\rangle \exp(-i\Delta t)\exp\left[-\left(\frac{J_{23}^2}{\Delta^2}\Gamma+\kappa\right)t\right]\\ &+\frac{J_{12}\Delta^2}{J^2\sqrt{J^2+\Delta^2}}\langle 3|\psi_3\rangle \exp\left[-\left(\frac{J_{23}^2}{J^2}\Gamma+\kappa\right)t\right],\\ =&-\frac{J_{12}J_{23}}{J^2}\exp\left[-\left(\frac{J_{12}^2}{J^2}\Gamma+\kappa\right)t\right]+\frac{J_{12}J_{23}}{J^2+\Delta^2}\exp(-i\Delta t)\exp\left[-\left(\frac{J_{23}^2}{\Delta^2}\Gamma+\kappa\right)t\right]\\ &+\frac{J_{12}J_{23}\Delta^2}{J^2(J^2+\Delta^2)}\exp\left[-\left(\frac{J_{23}^2}{J^2}\Gamma+\kappa\right)t\right], \hspace{2cm} (S84)\end{aligned}$$

$$\begin{aligned}\langle 3|\varphi_2(t)\rangle =&\frac{\Delta}{\sqrt{J^2+\Delta^2}}\left\{\langle 3|\psi_2\rangle \exp(-i\Delta t)\exp\left[-\left(\frac{J_{23}^2}{\Delta^2}\Gamma+\kappa\right)t\right]-\langle 3|\psi_3\rangle \exp\left[-\left(\frac{J_{23}^2}{J^2}\Gamma+\kappa\right)t\right]\right\},\\ =&\frac{J_{23}\Delta}{J^2+\Delta^2}\left\{\exp(-i\Delta t)\exp\left[-\left(\frac{J_{23}^2}{\Delta^2}\Gamma+\kappa\right)t\right]-\exp\left[-\left(\frac{J_{23}^2}{J^2}\Gamma+\kappa\right)t\right]\right\}, \hspace{1cm} (S85)\end{aligned}$$

$$\begin{aligned}\langle 3|\varphi_3(t)\rangle =&\frac{J_{12}}{J}\langle 3|\psi_1\rangle \exp\left[-\left(\frac{J_{12}^2}{J^2}\Gamma+\kappa\right)t\right]+\frac{J_{23}}{\sqrt{J^2+\Delta^2}}\langle 3|\psi_2\rangle \exp(-i\Delta t)\exp\left[-\left(\frac{J_{23}^2}{\Delta^2}\Gamma+\kappa\right)t\right]\\ &+\frac{J_{23}\Delta^2}{J^2\sqrt{J^2+\Delta^2}}\langle 3|\psi_3\rangle \exp\left[-\left(\frac{J_{23}^2}{J^2}\Gamma+\kappa\right)t\right],\\ =&\frac{J_{12}^2}{J^2}\exp\left[-\left(\frac{J_{12}^2}{J^2}\Gamma+\kappa\right)t\right]+\frac{J_{23}^2}{J^2+\Delta^2}\exp(-i\Delta t)\exp\left[-\left(\frac{J_{23}^2}{\Delta^2}\Gamma+\kappa\right)t\right]\\ &+\frac{J_{23}^2\Delta^2}{J^2(J^2+\Delta^2)}\exp\left[-\left(\frac{J_{23}^2}{J^2}\Gamma+\kappa\right)t\right]. \hspace{2cm} (S86)\end{aligned}$$

To the first order of $J/\Delta$, Eqs. (S84)-(S86) become

$$\langle 3|\varphi_1(t)\rangle =\frac{J_{12}J_{23}}{J^2}\left\{\exp\left[-\left(\frac{J_{23}^2}{J^2}\Gamma+\kappa\right)t\right]-\exp\left[-\left(\frac{J_{12}^2}{J^2}\Gamma+\kappa\right)t\right]\right\}, \hspace{1cm} (S87)$$

$$\langle 3|\varphi_2(t)\rangle =\frac{J_{23}}{\Delta}\left\{e^{-i\Delta t}\exp\left[-\left(\frac{J_{23}^2}{\Delta^2}\Gamma+\kappa\right)t\right]-\exp\left[-\left(\frac{J_{23}^2}{J^2}\Gamma+\kappa\right)t\right]\right\}, \hspace{1cm} (S88)$$

$$\langle 3|\varphi_3(t)\rangle =\frac{J_{12}^2}{J^2}\exp\left[-\left(\frac{J_{12}^2}{J^2}\Gamma+\kappa\right)t\right]+\frac{J_{23}^2}{J^2}\exp\left[-\left(\frac{J_{23}^2}{J^2}\Gamma+\kappa\right)t\right]. \hspace{1cm} (S89)$$

When the probabilities of the initial states being $|j\rangle$ ($j=1,2,3$) are equal, the population on the state $|3\rangle$ at time $t$ is

$$\begin{aligned}P_3(t)=&\frac{1}{3}\left[|\langle 3|\varphi_1(t)\rangle|^2+|\langle 3|\varphi_2(t)\rangle|^2+|\langle 3|\varphi_3(t)\rangle|^2\right],\\ =&\frac{1}{3}\left\{\frac{J_{12}^2}{J^2}\exp\left[-2\left(\frac{J_{12}^2}{J^2}\Gamma+\kappa\right)t\right]+\frac{J_{23}^2}{J^2}\exp\left[-2\left(\frac{J_{23}^2}{J^2}\Gamma+\kappa\right)t\right]\right.\\ &\left.+\frac{J_{23}^2}{\Delta^2}\left[\exp\left[-2\left(\frac{J_{23}^2}{\Delta^2}\Gamma+\kappa\right)t\right]+\exp\left[-2\left(\frac{J_{23}^2}{J^2}\Gamma+\kappa\right)t\right]+2\cos(\Delta t)\exp\left[-\left(\frac{J_{23}^2}{\Delta^2}\Gamma+\frac{J_{23}^2}{J^2}\Gamma+2\kappa\right)t\right]\right]\right\}\end{aligned}$$

The efficiency $\eta$ of this 3-level-system is

$$\begin{aligned}
\eta =& 2\Gamma \int_0^\infty \langle 3|\rho(t)|3\rangle \mathrm{d}t, \\
=& 2\Gamma \int_0^\infty P_3(t)\mathrm{d}t, \\
=& \frac{\Gamma}{3}\left[\frac{J_{12}^2}{J_{12}^2\Gamma + J^2\kappa} + \frac{J_{23}^2}{J_{23}^2\Gamma + J^2\kappa} + \frac{J_{23}^2}{J_{23}^2\Gamma + \Delta^2\kappa} + \frac{J_{23}^2 J^2}{\Delta^2(J_{23}^2\Gamma + J^2\kappa)} + \frac{J_{23}^2}{\Delta^4}\left(\frac{J_{23}^2}{\Delta^2}\Gamma + \frac{J_{23}^2}{J^2}\Gamma + 2\kappa\right)\right].
\end{aligned} \tag{S90}$$

We find that the last term of Eq. (S90) is much less than the others. Therefore, we could ignore this term and write the result as

$$\eta = \frac{\Gamma}{3}\left[\frac{J_{12}^2}{J_{12}^2\Gamma + J^2\kappa} + \frac{J_{23}^2}{J_{23}^2\Gamma + J^2\kappa} + \frac{J_{23}^2}{J_{23}^2\Gamma + \Delta^2\kappa} + \frac{J_{23}^2 J^2}{\Delta^2(J_{23}^2\Gamma + J^2\kappa)}\right]. \tag{S91}$$

The analytical result and the numerical calculation are compared in Fig. S3. As shown, the two result coincide with each other quite well in the whole parameter regime. And thus it is reasonable to obtain the efficiency the non-Hermitian Hamiltonian approach.

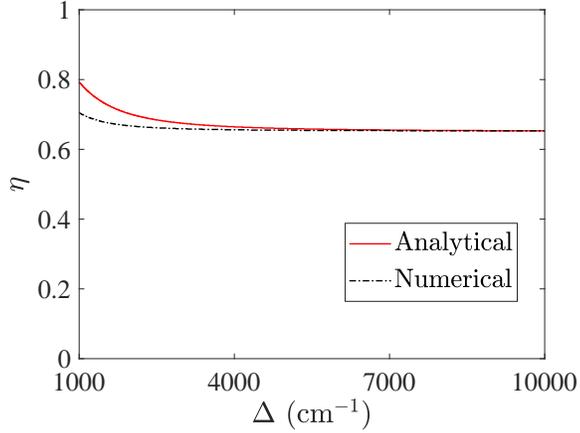

FIG. S3. Comparison of the analytical result and the numerical calculation. The other parameters are the same as in Fig. S2.

## II. THE DETAILED DERIVATION OF THE MASTER EQUATION

In this part, we demonstrate the detailed derivation of the master equation using the CMRT [4]. The Hamiltonian of the 'bridge' system is

$$H_S = \sum_{n=1}^{3} E_n |n\rangle\langle n| + \sum_{m \neq n} J_{mn} |m\rangle\langle n| = \sum_k \varepsilon'_k |\varepsilon_k\rangle\langle \varepsilon_k|. \tag{S92}$$

where $|n\rangle$ represents the excitation on the site $n$ with energy $E_n$ while the other sites on the ground state, $J_{mn}$ is the coupling between $|n\rangle$ and $|m\rangle$. The exciton state $|\varepsilon_k\rangle$ ($k = 1, 2, 3$) is a superposition of site bases as

$$|\varepsilon_k\rangle = \sum_n C_{nk} |n\rangle. \tag{S93}$$

The environment could be modeled as a collection of harmonic oscillators described by the Hamiltonian

$$\begin{aligned} H_E &= \sum_{n,q} \left[ \frac{p_{nq}^2}{2m_q} + \frac{1}{2} m_q \omega_q^2 x_{nq}^2 \right], \\ &= \sum_{n,q} \omega_q a_{nq}^\dagger a_{nq}, \end{aligned} \tag{S94}$$

where $a_{nq}^\dagger$ is the creation operator of $q$th harmonic oscillator of site $n$ with frequency $\omega_q$, and $x_{nq}, p_{nq}$ are the coordinate operator and the momentum operator of the site $n$. They satisfy

$$x_{nq} = \sqrt{\frac{1}{2m_q \omega_q}} (a_{nq}^\dagger + a_{nq}), \tag{S95}$$

$$p_{nq} = i \sqrt{\frac{m_q \omega_q}{2}} (a_{nq}^\dagger - a_{nq}). \tag{S96}$$

Thus, the interaction between the system and the environment is described by the Hamiltonian

$$\begin{aligned} H_{SE} &= \sum_{n,q} m_q \omega_q^2 x_{nq}^{(0)} x_{nq} |n\rangle\langle n|, \\ &= \sum_{n,q} g_{nq} \omega_q (a_{nq}^\dagger + a_{nq}) |n\rangle\langle n|, \end{aligned} \tag{S97}$$

where $g_{nq}$ is dimensionless coupling constant. By transformed to the eigen basis of $H_S$, the system-bath Hamiltonian reads

$$H_{SE} = \sum_{k,k',q} u_{kk'} |\varepsilon_k\rangle\langle \varepsilon_{k'}|, \tag{S98}$$

with the coupling coefficient

$$u_{kk'} = \sum_{n,q} C^*_{nk} C_{nk'} g_{nq} \omega_q (a^\dagger_{nq} + a_{nq}),$$
$$= \sum_{n,q} A_{kk'}(n) m_q \omega_q^2 x^{(0)}_{nq} x_{nq}, \tag{S99}$$

with

$$A_{kk'}(n) = C^*_{nk} C_{nk'}. \tag{S100}$$

The total Hamiltonian is the sum of these three part,i.e.,

$$H = H_S + H_E + H_{SE}. \tag{S101}$$

According to the CMRT [4, 5], the total Hamiltonian could be divided into the zeroth-order Hamiltonian including the diagonal system-bath interaction in the exciton basis, i.e.,

$$H_0 = \sum_k \left[\varepsilon'_k + u_{kk}\right] |\varepsilon_k\rangle\langle\varepsilon_k| + H_E, \tag{S102}$$

and a perturbation Hamiltonian, i.e., the off-diagonal system-bath interaction as

$$V = \sum_{k \neq k'} u_{kk'} |\varepsilon_k\rangle\langle\varepsilon_{k'}|. \tag{S103}$$

Noting that

$$\sum_n |n\rangle\langle n| = \sum_k |\varepsilon_k\rangle\langle\varepsilon_k| = 1, \tag{S104}$$

we rewrite $H_0$ as

$$\begin{aligned} H_0 &= \sum_{k,n,q} \left[\varepsilon'_k + A_{kk}(n) m_q \omega_q^2 x^{(0)}_{nq} x_{nq} + \left(\frac{p^2_{nq}}{2m_q} + \frac{1}{2} m_q \omega_q^2 x_{nq}^2\right)\right] |\varepsilon_k\rangle\langle\varepsilon_k| \\ &= \sum_k \left[\varepsilon_k + H_E(\{A_{kk}(n) x^{(0)}_{nq}\})\right] |\varepsilon_k\rangle\langle\varepsilon_k|, \end{aligned} \tag{S105}$$

where $\varepsilon_k$ and $H_E(\{A_{kk}(n) x^{(0)}_{nq}\})$ are respectively

$$\varepsilon_k = \varepsilon'_k - \Lambda_k, \tag{S106}$$
$$\Lambda_k = \frac{1}{2} A^2_{kk}(n) m_q \omega_q^2 (x^{(0)}_{nq})^2$$
$$= \sum_n A^2_{kk} \lambda_n, \tag{S107}$$
$$H_E(\{A_{kk}(n) x^{(0)}_{nq}\}) = \sum_{n,q} \left[\frac{p^2_{nq}}{2m_q} + \frac{1}{2} m_q \omega_q^2 (x_{nq} + A_{kk} x^{(0)}_{nq})^2\right]. \tag{S108}$$

where $\lambda_n$ is the reorganization energy for the exciton state $n$.

We define that $\chi(t)$ is the density operator for the total system including the system and the environment, and $\rho(t)$ is the reduced density matrix of the system, i.e.,

$$\rho(t) = \text{tr}_E[\chi(t)], \tag{S109}$$

where the trace is only taken over the environmental states. The von Neumann-Liouville equation for $\chi(t)$ reads

$$\dot{\chi}(t) = -i[H, \chi]. \tag{S110}$$

Transforming Eq. (S110) into the interaction picture with respect to $H_0$, we can obtain

$$\dot{\tilde{\chi}} = -i[\tilde{V}, \tilde{\chi}], \tag{S111}$$

with

$$\tilde{\chi}(t) = \exp(iH_0 t) \chi(t) \exp(-iH_0 t), \tag{S112}$$

$$\tilde{V}(t) = \exp(iH_0 t) V \exp(-iH_0 t). \tag{S113}$$

Integrating Eq. (S111) formally, we have

$$\tilde{\chi}(t) = \tilde{\chi}(0) - i \int_0^t [\tilde{V}(t'), \tilde{\chi}(t')] \mathrm{d}t'. \tag{S114}$$

Substituting the above equation into $\tilde{\chi}(t)$ on r.h.s of Eq. (S111) yields

$$\dot{\tilde{\chi}} = -i[\tilde{V}(t), \chi(0)] - \int_0^t [\tilde{V}(t), [\tilde{V}(t'), \tilde{\chi}(t')]] \mathrm{d}t'. \tag{S115}$$

In order to simplify the above equation, we will consider some reasonable approximations as follows. We assume that there is no initial correlation between the system and the environment. Thus, $\tilde{\chi}(0) = \chi(0)$ factorizes as

$$\chi(0) = \rho(0)\rho_E, \tag{S116}$$

where $\rho_E$ is the density operator of the environment at the initial stage. Then, noting that

$$\text{tr}_E\{\tilde{\chi}\} = \exp(iH_0 t) \rho \exp(-iH_0 t) = \tilde{\rho} \tag{S117}$$

After tracing over the environment, Eq. (S115) becomes

$$\dot{\tilde{\rho}} = -\int_0^t \text{tr}_E\{[\tilde{V}(t), [\tilde{V}(t'), \tilde{\chi}(t')]]\} \mathrm{d}t'. \tag{S118}$$

Here, we have neglected the term $-i\text{tr}_E\{[\tilde{V}(t), \tilde{\chi}(0)]\}$ under the assumption $\text{tr}_E[\tilde{V}(t)\rho_E] = 0$.

We have assumed that $\tilde{\chi}$ factorizes at $t = 0$. At later times the correlations between $S$ and $E$ will arise due to the coupling of the system and the environment via $V$. However, we have assumed that this coupling is weak. Thus, at all times, $\tilde{\chi}(t)$ should only show deviations of order $V$ from an uncorrelated state. Furthermore, $E$ is a large system whose state should be virtually unaffected by its coupling to $S$. Therefore, the density matrix of the total system reads

$$\tilde{\chi}(t) = \tilde{\rho}(t)\rho_E. \tag{S119}$$

By the perturbation theory, i.e., neglecting the terms higher than second order of $V$, Eq. (S118) becomes

$$\dot{\tilde{\rho}} = -\int_0^t \text{tr}_E\{[\tilde{V}(t), [\tilde{V}(t'), \tilde{\rho}(t')\rho_E]]\}dt'. \tag{S120}$$

We notice that Eq. (S120) is not Markovian since the time evolution of $\tilde{\rho}(t)$ depends on its past history through the integration over $\tilde{\rho}(t')$. Under the Markovian approximation, replacing $\tilde{\rho}(t')$ by $\tilde{\rho}(t)$ yields

$$\dot{\tilde{\rho}} = -\int_0^t \text{tr}_E\{[\tilde{V}(t), [\tilde{V}(t'), \tilde{\rho}(t)\rho_E]]\}dt'. \tag{S121}$$

Translating Eq. (S121) back to the Schrödinger picture, we obtain

$$\begin{aligned}
\dot{\rho} = & -i[H_0, \rho] - \int_0^t \text{tr}_E[V, [V(-\tau), \rho(t)\rho_E]]d\tau, \\
= & -i[H_0, \rho] - \int_0^t d\tau \text{tr}_E[VV(-\tau)\rho(t)\rho_E] + \int_0^t d\tau \text{tr}_E[V\rho(t)\rho_E V(-\tau)] + \int_0^t d\tau \text{tr}_E[V(-\tau)\rho(t)\rho_E V] \\
& - \int_0^t d\tau \text{tr}_E[\rho(t)\rho_E V(-\tau)V].
\end{aligned} \tag{S122}$$

where

$$\tau = t - t', \tag{S123}$$

$$V(-\tau) = \exp(-iH_0\tau)V\exp(iH_0\tau). \tag{S124}$$

### A. The dissipation rate

In this subsection, we deduce the dissipation rate for the diagonal elements of master equation, i.e., the the real coefficients of diagonal elements of density matrix on the right hand side of Eq. (S122).

The contribution of the second term is

$$-\int_0^t d\tau \langle \varepsilon_p | \text{tr}_E[VV(-\tau)\rho(t)\rho_E] | \varepsilon_p \rangle = -\int_0^t d\tau \text{tr}_E \left[ \sum_{k(\neq p)} \sum_{k'(\neq k)} u_{pk} e^{-iH_0^{(k)}\tau} u_{kk'} e^{iH_0^{(k')}\tau} \rho_{k'p} \rho_E \right],$$

$$\simeq -\int_0^t d\tau \text{tr}_E \left[ \sum_{k(\neq p)} u_{pk} e^{-iH_0^{(k)}\tau} u_{kp} e^{iH_0^{(p)}\tau} \rho_{pp} \rho_E \right]$$

$$= -\left\{ \int_0^t d\tau \langle \varepsilon_p | \text{tr}_E[\rho(t)\rho_E V(-\tau)V] | \varepsilon_p \rangle \right\}^*, \quad (S125)$$

where in the second line we have dropped the terms with $k' \neq p$, and

$$H_0^{(k)} = \varepsilon_k + H_E(a_{kk}(n)x_{nq}^{(0)}). \quad (S126)$$

In Eq. (S125) and the following, the symbols like $\rho_{kp}$ always represent

$$\rho_{kp} = \langle \varepsilon_k | \rho(t) | \varepsilon_p \rangle. \quad (S127)$$

Furthermore, the third term is given by

$$\int_0^t d\tau \text{tr}_E [\langle \varepsilon_p | V\rho(t)\rho_E V(\tau) ] | \varepsilon_p \rangle = \left\{ \int_0^t d\tau \text{tr}_E [\langle \varepsilon_p | V(-\tau)\rho(t)\rho_E V] | \varepsilon_p \rangle \right\},$$

$$= \int_0^t d\tau \text{tr}_E \left[ \sum_{k(\neq p)} \sum_{k'(\neq p)} u_{pk} \rho_{kk'} \rho_E e^{-iH_0^{(k)}\tau} u_{k'p} e^{iH_0^{(p)}\tau} \right],$$

$$\simeq \int_0^t d\tau \text{tr}_E \left[ \sum_{k(\neq p)} u_{pk} \rho_{kk} \rho_E e^{-iH_0^{(k)}\tau} u_{kp} e^{iH_0^{(p)}\tau} \right]. \quad (S128)$$

where in the last line we have dropped the terms with $k' \neq k$. Therefore, the equations of motion for diagonal elements are

$$\dot{\rho}_{k'k'} = -i\langle \varepsilon_{k'} | [H_0, \rho] | \varepsilon_{k'} \rangle - \sum_{k(\neq k')} R_{kk'}^{dis}(t) \rho_{k'k'} + \sum_{k(\neq k')} R_{k'k}^{dis}(t) \rho_{kk} \quad (S129)$$

with the dissipation rate from $|\varepsilon_{k'}\rangle$ to $|\varepsilon_k\rangle$ defined by

$$R_{kk'}^{dis}(t) = 2\text{Re} \int_0^t d\tau \text{tr}_E \left[ u_{kk'} e^{-iH_0^{(k)}\tau} u_{k'k} e^{iH_0^{(k')}\tau} \rho_E \right]. \quad (S130)$$

Here, the density matrix of the bath $\rho_E$ is assumed to be $\exp(-\beta H_0^{(k')})$ corresponding to the initial state of the exciton $|\varepsilon_{k'}\rangle$, where

$$\beta = \frac{1}{k_B T}. \quad (S131)$$

In order to obtain the explicit expression of $R^{dis}_{kk'}(t)$, we introduce the displacement operator as

$$D(\{x^{(0)}_{nq}\}) = \prod_{n,q} D_{nq}(x^{(0)}_{nq}), \tag{S132}$$

$$\begin{aligned} D_{nq}(x^{(0)}_{nq}) &= \exp(-ix^{(0)}_{nq}p_{nq}) \\ &= \exp\left[\frac{1}{2}d^{(k)}_{nq}(a^\dagger_{nq} - a_{nq})\right] \\ &= D(\frac{1}{2}d^{(k)}_{nq}), \end{aligned} \tag{S133}$$

where in the second line we use the result of Eq. (S96). The displaced Hamiltonian is written as

$$H_E(\{\frac{1}{2}d^{(k)}_{nq}\}) = D^\dagger(\{\frac{1}{2}d^{(k)}_{nq}\}) H_E D(\{\frac{1}{2}d^{(k)}_{nq}\}), \tag{S134}$$

where the displacement is

$$d^{(k)}_{nq} = A_{kk}(n) d_{nq}, \tag{S135}$$

$$d_{nq} = \sqrt{2m_q\omega_q} x^{(0)}_{nq}. \tag{S136}$$

At this stage, we define a generating function as

$$F(y,z;\tau) = \prod_{n,q} F_{nq}(y,z;\tau), \tag{S137}$$

with

$$F_{nq}(y,z;\tau) = \text{tr}_E\left[e^{iH_E^{(nq)}(\frac{1}{2}d^{(k')}_{nq})\tau} e^{yu_{nq}A_{k'k}(n)} e^{-iH_E^{(nq)}(\frac{1}{2}d^{(k)}_{nq})\tau} e^{zu_{nq}A_{kk'}(n)} e^{-\beta H_E^{(nq)}(\frac{1}{2}d^{(k')}_{nq})}\right], \tag{S138}$$

$$u_{nq} = m_q\omega_q^2 x^{(0)}_{nq} x_{nq} = \frac{1}{2}\omega_q d_{nq}(a^\dagger_{nq} + a_{nq}). \tag{S139}$$

By repeatedly making use of the Baker-Hausdorff formula

$$e^A e^B = e^{A+B} e^{[A,B]/2}, \tag{S140}$$

for

$$[[A,B],A] = [[A,B],B] = 0, \tag{S141}$$

we have

$$\begin{aligned} &F_{nq}(y,z;\tau) \\ &= \text{tr}_E\left[e^{iH_E^{(nq)}\tau} D_{nq}(\frac{1}{2}d^{(k')}_{nq}) e^{yu_{nq}A_{k'k}(n)} D^\dagger_{nq}(\frac{1}{2}d^{(k)}_{nq}) e^{-iH_E^{(nq)}\tau} D_{nq}(\frac{1}{2}d^{(k)}_{nq}) e^{zu_{nq}A_{kk'}(n)} D^\dagger_{nq}(\frac{1}{2}d^{(k')}_{nq}) e^{-\beta H_E^{(nq)}}\right], \\ &= \text{tr}_E\left[e^{(A_1+B_1+C_1)a^\dagger_{nq}(t) - (A_1-B_1+C_1)a_{nq}(t)} e^{(A_2+B_2+C_2)a^\dagger_{nq} - (A_2-B_2+C_2)a_{nq}} e^{-\beta\omega_q a^\dagger_{nq} a_{nq}}\right] \\ &\times e^{-(A_1-C_1)B_1} e^{-(A_2-B_2)C_2}, \\ &= \exp\left[\frac{1}{2}(M_1+N_1)(M_2+N_2)\coth\left(\frac{\beta\omega_q}{2}\right)\right] \exp(M_3+N_3), \end{aligned} \tag{S142}$$

where

$$M_1 = (A_1 + B_1 + C_1)e^{i\omega_q \tau}, \tag{S143}$$

$$M_2 = -(A_1 - B_1 + C_1)e^{-i\omega_q \tau}, \tag{S144}$$

$$M_3 = -(A_1 - C_1)B_1 - (A_2 - C_2)B_2, \tag{S145}$$

$$N_1 = A_2 + B_2 + C_2, \tag{S146}$$

$$N_2 = -(A_2 - B_2 + C_2), \tag{S147}$$

$$N_3 = \frac{1}{2}(-M_1 N_2 + M_2 N_1), \tag{S148}$$

$$A_1 = \frac{1}{2} d_{nq} A_{k'k'}(n), \tag{S149}$$

$$B_1 = \frac{1}{2} d_{nq} A_{k'k}(n) y \omega_q, \tag{S150}$$

$$C_1 = -\frac{1}{2} d_{nq} A_{kk}(n), \tag{S151}$$

$$A_2 = \frac{1}{2} d_{nq} A_{kk'}(n), \tag{S152}$$

$$B_2 = \frac{1}{2} d_{nq} A_{kk'}(n) z \omega_q, \tag{S153}$$

$$C_2 = -\frac{1}{2} d_{nq} A_{kk'}(n). \tag{S154}$$

and we have used the relation

$$e^{iH_E^{(nq)}(\{\frac{1}{2}d_{nq}^{(k)}\})\tau} = D_{nq}^\dagger(\frac{1}{2}d_{nq}^{(k)})e^{iH_E^{(nq)}\tau}D_{nq}(\frac{1}{2}d_{nq}^{(k)}), \tag{S155}$$

$$\mathrm{tr}_E\left[e^{r_1 a_{nq} + r_2 a_{nq}^\dagger} e^{-\beta \omega_q a_{nq}^\dagger a_{nq}}\right] = e^{\frac{1}{2}r_1 r_2 \coth\left(\frac{\beta \omega_q}{2}\right)}, \tag{S156}$$

$$e^{iH_E^{(nq)}\tau} a_{nq}^\dagger e^{-iH_E^{(nq)}\tau} = a_{nq}^\dagger e^{i\omega_q \tau}. \tag{S157}$$

Replacing the summation over $q$ by the integral over frequency $\omega$, i.e.,

$$\sum_q \frac{1}{4} d_{nq}^2 f(\omega_q) = \int f(\omega) \frac{J_n(\omega)}{\omega^2} d\omega, \tag{S158}$$

with $J_n(\omega)$ being the interacting spectrum of the nth site, we can rewrite the generating function as

$$\prod_{n,q} F_{nq}(y, z; \tau) = \exp\left[\sum_n (c_5^{(n)} y^2 + c_4^{(n)} z^2 + c_3^{(n)} yz + c_2^{(n)} y + c_1^{(n)} z + c_0^{(n)})\right], \tag{S159}$$

where the corresponding coefficients to be used for calculating rates are

$$c_0^{(n)} = -[A_{kk}(n) - A_{k'k'}(n)]^2[g_n(\tau) + i\lambda_n\tau], \tag{S160}$$

$$c_1^{(n)} = -iA_{k'k}(n)[A_{kk}(n) - A_{k'k'}(n)]\dot{g}_n(\tau) - 2A_{k'k}(n)A_{k'k'}(n)\lambda_n, \tag{S161}$$

$$c_2^{(n)} = -iA_{kk'}(n)[A_{kk}(n) - A_{k'k'}(n)]\dot{g}_n(\tau) - 2A_{kk'}(n)A_{k'k'}(n)\lambda_n, \tag{S162}$$

$$c_3^{(n)} = A_{k'k}(n)A_{kk'}(n)\ddot{g}_n(\tau). \tag{S163}$$

Since the dissipation rate is given by

$$R_{kk'}^{dis}(t) = 2\text{Re}\int_0^t d\tau e^{-i(\varepsilon_k-\varepsilon_{k'})\tau}\lim_{y,z\to 0}\frac{\partial^2}{\partial y\partial z}F(y,z;\tau)$$
$$= 2\text{Re}\int_0^t d\tau e^{-i(\varepsilon_k-\varepsilon_{k'})\tau}\left[\sum_n c_3^{(n)} + \sum_n c_1^{(n)}c_2^{(n)}\right]\exp\left[\sum_n c_0^{(n)}\right], \tag{S164}$$

we finally obtain

$$R_{kk'}^{dis}(t) = 2\text{Re}\int_0^t d\tau e^{-i(\varepsilon_k-\varepsilon_{k'})\tau}e^{-[g_{kkkk}(\tau)+g_{k'k'k'k'}(\tau)-2g_{kkk'k'}(\tau)]}e^{-i(\Lambda_k+\Lambda_{k'}-2\lambda_{kkk'k'})\tau}$$
$$\times\{\ddot{g}_{k'kkk'}(\tau) - [\dot{g}_{k'kkk}(\tau) - \dot{g}_{k'kk'k'}(\tau) - 2i\lambda_{k'kk'k'}][\dot{g}_{kk'kk}(\tau) - \dot{g}_{kk'k'k'}(\tau) - 2i\lambda_{kk'k'k'}]\}, \tag{S165}$$

where

$$g_{k_1k_2k_3k_4}(t) = \sum_n A_{k_1k_2}(n)A_{k_3k_4}(n)g_n(t), \tag{S166}$$

$$\lambda_{k_1k_2k_3k_4} = \sum_n A_{k_1k_2}(n)A_{k_3k_4}(n)\lambda_n, \tag{S167}$$

$$g_n(t) = \int_0^\infty d\omega \frac{J_n(\omega)}{\omega^2}\left[(1-\cos\omega t)\coth\left(\frac{\beta\omega}{2}\right) + i(\sin\omega t - \omega t)\right]. \tag{S168}$$

**B. The pure dephasing rate**

In this section, we will derive the pure-dephasing rate for the master equation, i.e., the real coefficients of off-diagonal elements of density matrix on the right hand side of Eq. (S122). Since the density matrix of the total system involving the system and the environment is

$$\chi(t) = e^{-iHt}\chi(0)e^{iHt}, \tag{S169}$$

the off-diagonal term of the reduced density matrix for the system is

$$\begin{aligned}
\rho_{kk'}(t) &= \mathrm{tr}_E\left[\langle\varepsilon_k|\mathrm{e}^{-iHt}\chi(0)\mathrm{e}^{iHt}|\varepsilon_{k'}\rangle\right] \\
&\simeq c_{kk'}(0)\mathrm{tr}_E\left[\langle\varepsilon_k|\mathrm{e}^{-iH_0t}|\varepsilon_k\rangle\langle\varepsilon_{k'}|\rho_E\mathrm{e}^{iH_0t}|\varepsilon_{k'}\rangle\right] \\
&= c_{kk'}(0)\mathrm{tr}_E\left[\mathrm{e}^{-iH_0^{(k)}t}\mathrm{e}^{-\beta H_E(\{\frac{1}{2}d_{nq}^{(k')}\})}\mathrm{e}^{iH_0^{(k')}t}\right] \\
&= c_{kk'}(0)\mathrm{e}^{-i(\varepsilon_k-\varepsilon_{k'})t}\mathrm{tr}_E\left[\mathrm{e}^{-iH_E(\{\frac{1}{2}d_{nq}^{(k)}\})t}\mathrm{e}^{-\beta H_E(\{\frac{1}{2}d_{nq}^{(k')}\})}\mathrm{e}^{iH_E(\{\frac{1}{2}d_{nq}^{(k')}\})t}\right] \\
&= c_{kk'}(0)\mathrm{e}^{-i(\varepsilon_k-\varepsilon_{k'})t}\prod_{n,q}\mathrm{tr}_E\left[D_{nq}(-\frac{1}{2}d_{nq}^{(k)})\mathrm{e}^{-iH_Et}D_{nq}(\frac{1}{2}d_{nq}^{(k)})D_{nq}(-\frac{1}{2}d_{nq}^{(k')})\mathrm{e}^{-\beta H_E}\mathrm{e}^{-iH_Et}D_{nq}(\frac{1}{2}d_{nq}^{(k')})\right] \\
&= c_{kk'}(0)\mathrm{e}^{-i(\varepsilon_k-\varepsilon_{k'})t}\prod_{n,q}\mathrm{tr}_E\left[D_{nq}\left(\frac{1}{2}d_{nq}^{(k')}-\frac{1}{2}d_{nq}^{(k)}\right)\mathrm{e}^{-iH_Et}D_{nq}\left(\frac{1}{2}d_{nq}^{(k)}-\frac{1}{2}d_{nq}^{(k')}\right)\mathrm{e}^{iH_Et}\mathrm{e}^{-\beta H_E}\right] \\
&= c_{kk'}(0)\mathrm{e}^{-i(\varepsilon_k-\varepsilon_{k'})t}\prod_{n,q}\mathrm{tr}_E\left[D_{nq}\left(\frac{1}{2}d_{nq}^{(k')}-\frac{1}{2}d_{nq}^{(k)}\right)D_{nq}\left(\left(\frac{1}{2}d_{nq}^{(k)}-\frac{1}{2}d_{nq}^{(k')}\right)\mathrm{e}^{-i\omega_q t}\right)\mathrm{e}^{-\beta H_E}\right] \\
&= c_{kk'}(0)\mathrm{e}^{-i(\varepsilon_k-\varepsilon_{k'})t}\prod_{n,q}\mathrm{tr}_E\left[D_{nq}\left(\left(\frac{1}{2}d_{nq}^{(k)}-\frac{1}{2}d_{nq}^{(k')}\right)(\mathrm{e}^{-i\omega_q t}-1)\right)\mathrm{e}^{-\beta H_E}\right] \\
&= c_{kk'}(0)\mathrm{e}^{-i(\varepsilon_k-\varepsilon_{k'})t}\prod_{n,q}\mathrm{tr}_E\left[\exp\left(\left(\frac{1}{2}d_{nq}^{(k)}-\frac{1}{2}d_{nq}^{(k')}\right)(\mathrm{e}^{-i\omega_q t}-1)a_{nq}^\dagger - \mathrm{h.c.}\right)\mathrm{e}^{-\beta\omega_q a_{nq}^\dagger a_{nq}}\right] \\
&= c_{kk'}(0)\mathrm{e}^{-i(\varepsilon_k-\varepsilon_{k'})t}\prod_{n,q}\exp\left[-\frac{1}{2}\left(\frac{1}{2}d_{nq}^{(k)}-\frac{1}{2}d_{nq}^{(k')}\right)^2|\mathrm{e}^{-i\omega_q t}-1|^2\coth\left(\frac{\beta\omega_q}{2}\right)\right] \\
&= c_{kk'}(0)\mathrm{e}^{-i(\varepsilon_k-\varepsilon_{k'})t}\exp\left[-\sum_n(A_{kk}(n)-A_{k'k'}(n))^2\int_0^\infty \mathrm{d}\omega\frac{J_n(\omega)}{\omega^2}(1-\cos\omega t)\coth\left(\frac{\beta\omega}{2}\right)\right].
\end{aligned}$$
(S170)

And its derivative with respect to the time is

$$\dot{\rho}_{kk'} = -i(\varepsilon_k-\varepsilon_{k'})\rho_{kk'} - R_{kk'}^{pd}(t)\rho_{kk'}, \tag{S171}$$

where the pure-dephasing rate is

$$\begin{aligned}
R_{kk'}^{pd}(t) &= \sum_n[A_{kk}(n)-A_{k'k'}(n)]^2\int_0^\infty \mathrm{d}\omega\frac{J_n(\omega)}{\omega}\sin\omega t\coth\left(\frac{\beta\omega}{2}\right), \\
&= \sum_n[A_{kk}(n)-A_{k'k'}(n)]^2\mathrm{Re}[\dot{g}_n(t)].
\end{aligned} \tag{S172}$$

In the above derivation, we have used Eq. (S156) and the properties of the displacement operator

$$D(\alpha)D(\beta) = D(\alpha+\beta), \tag{S173}$$

$$D^\dagger(\alpha) = D(-\alpha). \tag{S174}$$

All in all, on account of the effects of dissipation and pure-dephasing, we obtain the complete master equation

$$\partial_t \rho = -i[H_S(t), \rho] - \sum_{k \neq k'} R^{dis}_{kk'}(t) \left[ \{\mathcal{A}^\dagger_{kk'} \mathcal{A}_{kk'}, \rho\} - 2\mathcal{A}_{kk'} \rho \mathcal{A}^\dagger_{kk'} \right] - \sum_{k \neq k'} R^{pd}_{kk'}(t) \rho_{kk'} \mathcal{A}_{kk'}, \quad \text{(S175)}$$

where $\{\mathcal{A}^\dagger_{kk'} \mathcal{A}_{kk'}, \rho\}$ are the anti-commutator and the jumping operators $\mathcal{A}_{kk'}$ are

$$\mathcal{A}_{kk'} = |\varepsilon_k\rangle\langle\varepsilon_{k'}|. \quad \text{(S176)}$$

For further simplification, we could rewrite the pure-dephasing term in the Lindblad form. Noticing that the jumping operator $\mathcal{A}_{kk}$ satisfy

$$\mathcal{A}^\dagger_{kk} \mathcal{A}_{kk} = \mathcal{A}_{kk} = \mathcal{A}^\dagger_{kk} = |\varepsilon_k\rangle\langle\varepsilon_k|, \quad \text{(S177)}$$

we have

$$\begin{aligned}
\{\mathcal{A}^\dagger_{kk} \mathcal{A}_{kk}, \rho\} - 2\mathcal{A}_{kk} \rho \mathcal{A}^\dagger_{kk} &= |\varepsilon_k\rangle\langle\varepsilon_k|\rho + \rho|\varepsilon_k\rangle\langle\varepsilon_k| - 2|\varepsilon_k\rangle\langle\varepsilon_k|\rho|\varepsilon_k\rangle\langle\varepsilon_k|, \\
&= |\varepsilon_k\rangle\langle\varepsilon_k|\rho + \rho|\varepsilon_k\rangle\langle\varepsilon_k| - 2\mathcal{A}_{kk}\rho_{kk}, \\
&= \sum_{k'} (\mathcal{A}_{kk'} \rho_{kk'} + \mathcal{A}_{k'k} \rho_{k'k}) - 2\mathcal{A}_{kk}\rho_{kk}, \\
&= \sum_{k'(\neq k)} (\mathcal{A}_{kk'} \rho_{kk'} + \mathcal{A}_{k'k} \rho_{k'k}).
\end{aligned} \quad \text{(S178)}$$

Furthermore,

$$\begin{aligned}
-\sum_k \frac{\Gamma_k}{2} \left[ \{\mathcal{A}^\dagger_{kk} \mathcal{A}_{kk}, \rho\} - 2\mathcal{A}_{kk} \rho \mathcal{A}^\dagger_{kk} \right] &= -\sum_k \sum_{k'(\neq k)} \frac{\Gamma_k}{2} (\mathcal{A}_{kk'} \rho_{kk'} + \mathcal{A}_{k'k} \rho_{k'k}), \\
&= -\sum_k \sum_{k'(\neq k)} \left( \frac{\Gamma_k}{2} \mathcal{A}_{kk'} \rho_{kk'} + \frac{\Gamma_{k'}}{2} \mathcal{A}_{kk'} \rho_{kk'} \right), \\
&= -\sum_k \sum_{k'(\neq k)} \frac{\Gamma_k + \Gamma_{k'}}{2} \mathcal{A}_{kk'} \rho_{kk'},
\end{aligned} \quad \text{(S179)}$$

where in the second term of the second line we exchange the subscripts $k$ and $k'$. Therefore, if we could rewrite the pure-dephasing rate $R^{pd}_{kk'}$ as

$$R^{pd}_{kk'}(t) = \frac{\Gamma_k + \Gamma_{k'}}{2}, \quad \text{(S180)}$$

the pure-dephasing term can become the Lindblad form

$$-\sum_{k \neq k'} R^{pd}_{kk'}(t) \rho_{kk'} \mathcal{A}_{kk'} = -\sum_k \frac{\Gamma_k}{2} \left[ \{\mathcal{A}^\dagger_{kk} \mathcal{A}_{kk}, \rho\} - 2\mathcal{A}_{kk} \rho \mathcal{A}^\dagger_{kk} \right]. \quad \text{(S181)}$$

Assuming that the total number of the sites is $N$, there are $N(N-1)/2$ independent pure-dephasing rates $R^{pd}_{kk'}$ but only $N$ Lindblad form pure-dephasing rates $\Gamma_k$. Thus we need some additional constraints to determine $\Gamma_{k'}$. Here we adopt a square-least fit method to obtain the final results. According to Ref. [4], the fitting results are in good agreement with the experimental results. Therefore, we require the mean-square displacement to be minimal with respect to all the Lindblad rates:

$$\frac{\partial}{\partial \Gamma_a} \sum_{k=1}^{N-1} \sum_{k'=k+1}^{N} \left[ R^{pd}_{kk'} - \frac{1}{2}(\Gamma_k + \Gamma_{k'}) \right]^2 = 0. \tag{S182}$$

where $a = 1, 2, \ldots, N$. After some mathematical derivation, we obtain a system of linear equations for the Lindblad-form dephasing rates, that is

$$\frac{1}{2}\sum_{k=1}^{a-1} \Gamma_k + \frac{1}{2}(2N-a)\Gamma_a + \sum_{k=a+1}^{N-1} \Gamma_k + \frac{1}{2}\Gamma_N = P_a, \tag{S183}$$

where the coefficient on the right hand side is

$$P_a = \sum_{k=a+1}^{N} R^{pd}_{ak} + \sum_{k=1}^{N-1} R^{pd}_{ka}. \tag{S184}$$

Equivalently, Eq. (S183) can be rewritten in a matrix form as

$$Q\Gamma = P, \tag{S185}$$

with the matrix elements of $Q$ given by

$$Q_{jk} = \begin{cases} \frac{1}{2}, & \text{for } k < j, \\ \frac{1}{2}(2N-j), & \text{for } k = j, \\ 1, & \text{for } j < k < N, \\ \frac{1}{2}, & \text{otherwise.} \end{cases} \tag{S186}$$

Therefore, by multiplying both sides with the inverse of $Q$, we have

$$\Gamma = Q^{-1}P. \tag{S187}$$

Finally, the original master equation Eq. (S175) can be rewritten in the Lindblad form as

$$\partial_t \rho = -i[H_S(t), \rho] - \sum_{k,k'} R_{kk'}(t) \left[ \left\{ \mathcal{A}^{\dagger}_{kk'} \mathcal{A}_{kk'}, \rho \right\} - 2\mathcal{A}_{kk'} \rho \mathcal{A}^{\dagger}_{kk'} \right], \tag{S188}$$

where the matrix elements of the population transfer and dephasing rates $R_{kk'}$ are defined respectively as

$$R_{kk'} \equiv \begin{cases} R_{kk'}^{dis}(t), & \text{for } k \neq k', \\ \dfrac{\Gamma_k}{2}, & \text{for } k = k'. \end{cases} \quad \text{(S189)}$$

In the theoretical derivation above, the upper limits of all integrals have been replaced by infinite. This operation is equivalent to making the Markovian approximation to our theory. Since the non-Markovian effects are not significant in the energy transfer, i.e., the characteristic time of the environmental relaxation is much less than the duration of system evolution, it is reasonable to make the Markovian approximation and thus significantly simplify the calculation.


	
\end{document}